\documentclass[10pt,a4paper,twoside,dvipdfm]{article}
\usepackage{epsfig}
\usepackage{baltlat6}
\usepackage{array}
\usepackage{wrapfig}
\usepackage{longtable}
\usepackage{lscape}
\begin{document}
\ \
\vspace{0.5mm}
\setcounter{page}{327}
\vspace{8mm}

\titlehead{Baltic Astronomy, vol.\,16, 327--347, 2007}

\titleb{YOUNG STARS IN THE CAMELOPARDALIS DUST AND\\ MOLECULAR CLOUDS.
II. INFRARED OBJECTS}

\begin{authorl}
\authorb{V. Strai\v zys}{} and
\authorb{V. Laugalys}{}
\end{authorl}

\moveright-3.2mm
\vbox{
\begin{addressl}
\addressb{}{Institute of Theoretical Physics and Astronomy, Vilnius
University,\\ Go\v stauto 12, Vilnius LT-01108, Lithuania}
\end{addressl}}

\submitb{Received 2007 August 10; accepted 2007 September 20}

\begin{summary} Using infrared photometric data extracted from the
2MASS, IRAS and MSX databases, 142 suspected young stellar objects
(YSOs) are selected from about 2 million stars in the Camelopardalis
segment of the Milky Way limited by Galactic coordinates $\ell$, $b$
= 132--158\degr, $\pm$12\degr.  According to radial velocities of
the associated CO clouds, the objects are attributed to three molecular
and dust cloud layers at 150--300 pc, $\sim$\,900 pc and 2.2 kpc
distances from the Sun.  These objects concentrate into dust and
molecular clouds and exhibit extremely large reddenings ($A_V$ up to 25
mag) which can be caused by the dust in foreground clouds and
circumstellar envelopes or disks.  In the $J$--$H$~vs.~$H$--$K$ diagram
these objects lie above the intrinsic line of T Tauri variables, roughly
along the black-body line.  Among the identified objects, some already
known YSOs are present, including the well investigated massive object
GL\,490.  The spectral energy distributions between 700 nm and 100
$\mu$m suggest that the objects may be YSOs of classes I, II and III.
However, we do not exclude the possibility that a small fraction of the
objects, especially those without IRAS and MSX photometry, may be
unrecognized heavily reddened OB-stars, late-type AGB stars or even
galaxies.  \end{summary}


\resthead{Young stars in the Camelopardalis dust and molecular clouds.
II.}{V. Strai\v zys, V. Laugalys}

\sectionb{1}{INTRODUCTION}

In the previous paper (Strai\v zys \& Laugalys 2007, Paper I) we have
shown that the star-forming process in the Camelopardalis segment of the
Local spiral arm is still active.  This was testified by the presence in
the area of more than 40 stars of the Cam OB1 association and of about
20 young stars of lower masses exhibiting emission in H$\alpha$ or
belonging to irregular variable stars of types IN and IS.  A high-mass
young stellar object, GL 490, embedded in the densest part of the dust
cloud DoH 942 (Dobashi et al. 2005), has been identified by Snell et
al.  (1984) and investigated in many subsequent papers.

Trying to find more young stellar objects (hereafter YSOs) of different
masses in the area limited by the Galactic coordinates $\ell$, $b$ =
132--158\degr, $\pm$\,12\degr\ we have analyzed the infrared objects
measured in the 2MASS, IRAS and MSX surveys.

\sectionb{2}{IDENTIFICATION OF PRE-MAIN-SEQUENCE OBJECTS}

Figure 1 shows the $J$--$H$~vs.~$H$--$K_s$ diagram for about
2\,$\times$\,$10^6$ stars measured in the 2MASS survey with the errors
$\leq$\,0.05 mag (Cutri et al. 2003; Skrutskie et al. 2006).  In the
comet-like crowding of dots the orange line designates the intrinsic
main sequence, the yellow line K--M giants and the blue line
the intrinsic locus of T Tauri-type stars from Meyer et al.
(1997). The red line is the interstellar reddening vector; its length
corresponds to the extinction $A_V$ = 10 mag in the $V$ passband.  The
violet line shows the black-body locus.  The `comet head' is
composed mostly of normal stars of different spectral classes and
luminosities with little or no interstellar reddening.  The upper
rich `tail 1' of the `comet' is composed of normal reddened
background stars, mostly of K and M giants.  This `tail' extends up to
$H$--$K_s$\,$\approx$\,1.5, i.e., some of the stars exhibit an
extinction $A_V$ of 20--25 mag.


\begin{figure}[!t]
\centerline{\psfig{figure=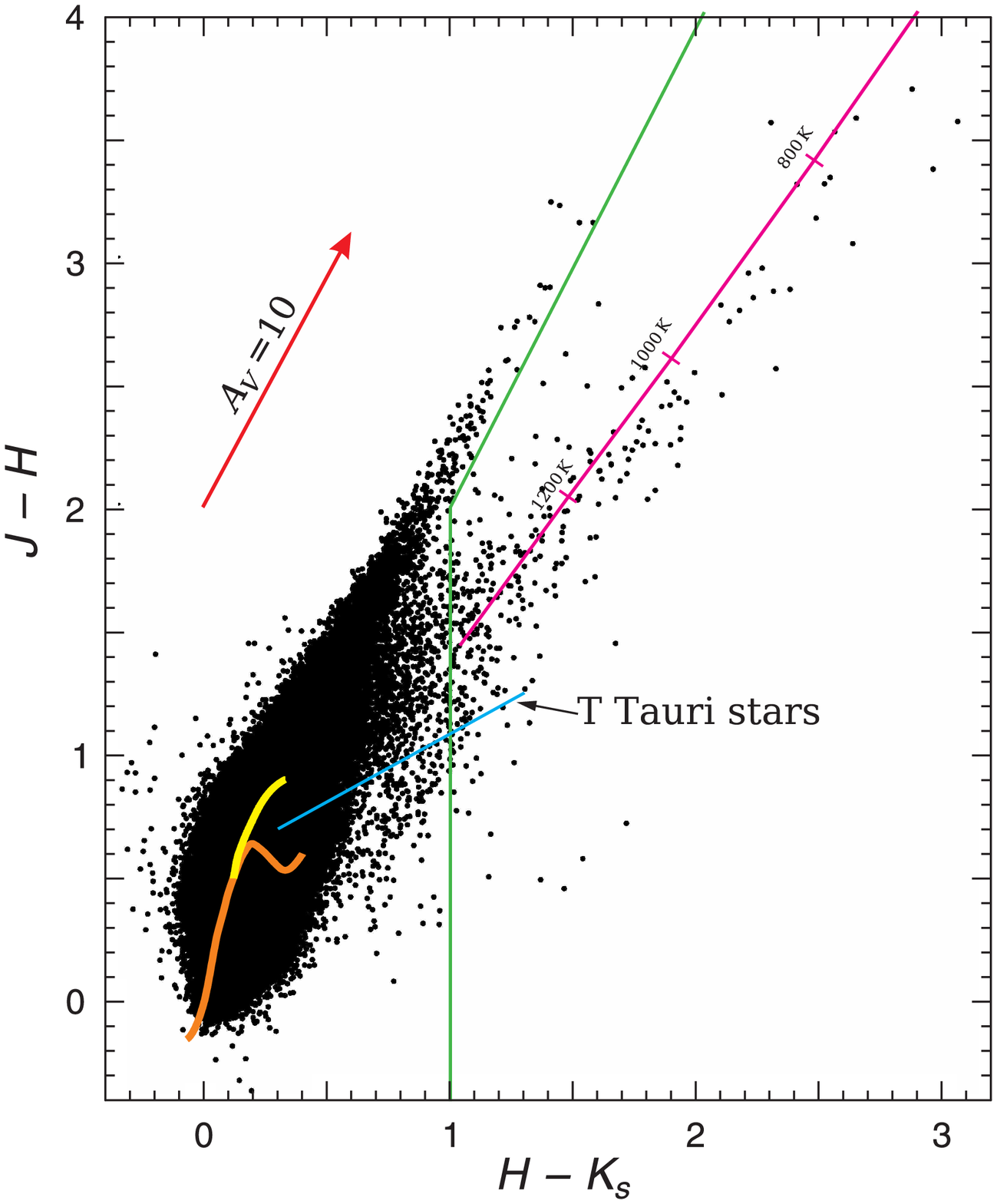,width=100mm,angle=0,clip=true}}
\vspace{0mm}
\captionb{1}{ The $J$--$H$ vs.  $H$--$K_s$ diagram for 2 million stars
in the investigated area.  The intrinsic main-sequence and K--M giant
lines are shown in orange and yellow, respectively.  The blue line
designates the intrinsic locus of T Tauri stars, the violet line is the
locus of black bodies.  The length of the reddening vector (shown in
red) corresponds to the extinction in the $V$ passband of 10 mag.  The
two green intersecting lines separate the region where the presence of
young stellar objects was investigated.}
\end{figure}

The lower tail (hereafter `tail 2') is much longer, reaching
$H$--$K_s$\,$\approx$\,3.0 and running more or less along the black-body
line.  In this part of the diagram we expect to find a variety of
reddened stars in early and late stages of evolution and extragalactic
objects:  M-type giants of the latest subclasses, including oxygen-rich
and carbon-rich long-period variables (Whitelock et al. 1994, 2000),
OH/IR stars (Sevenster 2002, Lewis et al. 2004, Jim\'enez-Esteban et al.
2005, 2006), carbon-rich stars of spectral type N (Whitelock et al.
2006), T Tauri-type stars with dense disks (Lada \& Adams 1992, Kenyon
\& Hartmann 1995), young stellar objects surrounded by gas and dust
envelopes (Persson \& Campbell 1987, 1988; Campbell et al. 1989), Ae/Be
stars (Lada \& Adams 1992, Hillenbrand et al. 1992), Be stars (Dougherty
et al. 1994), galaxies and quasars (Finlator et al. 2000, Ivezi\'c et
al. 2002).  Unfortunately, {\em JHK} photometry is not a sufficient tool
for the identification of young stars among such a variety of other
objects.  Therefore, either spectroscopic or infrared photometric
observations at longer wavelengths are essential.

For further analysis we isolated about 300 objects of `tail 2' by two
intersecting lines (green lines in Figure 1).  The lower vertical line
runs along $H$--$K_s$\,=\,1.0 and the upper line is the interstellar
reddening vector corresponding to
$$
Q_{JHK} = (J - H) - 1.85 (H - K_s) = 0.0.
$$
With this undertaking we have excluded the majority of normal stars of
various temperatures, luminosities and reddenings (except reddened O--B
stars and the coolest M giants and dwarfs).  The known M- and N-type
variables of the asymptotic giant branch (including the OH/IR objects),
Be stars, quasars and galaxies were identified in the available catalogs
and removed from the list.  About ten objects at Galactic latitudes
$>$\,5\degr, lying outside the dense interstellar clouds, were also
excluded.  Some galaxies were recognized and excluded by inspecting
their images in the SkyView database.  For the classification of the
remaining objects we have used color indices from the IRAS catalog of
point sources (Beichman et al. 1988).

The IRAS 12, 25 and 60 $\mu$m fluxes of reasonable accuracy ($\sigma$ of
the flux $<$\,20\,\%) are available for about 1/3 of the objects in the
remaining list.  For the identification of T Tauri stars and other YSOs,
the two-color diagram [12]--[25] vs.  [25]--[60] shown in Figure 2 was
used. Here, the IRAS color indices (as defined by Walker \& Cohen
1988) are:
$$
[12] - [25] = 1.56 - 2.5 \log F_{12}/F_{25},
$$
$$
[25] - [60] = 1.88 - 2.5 \log F_{25}/F_{60}.
$$


\begin{figure}[!t]
\centerline{\psfig{figure=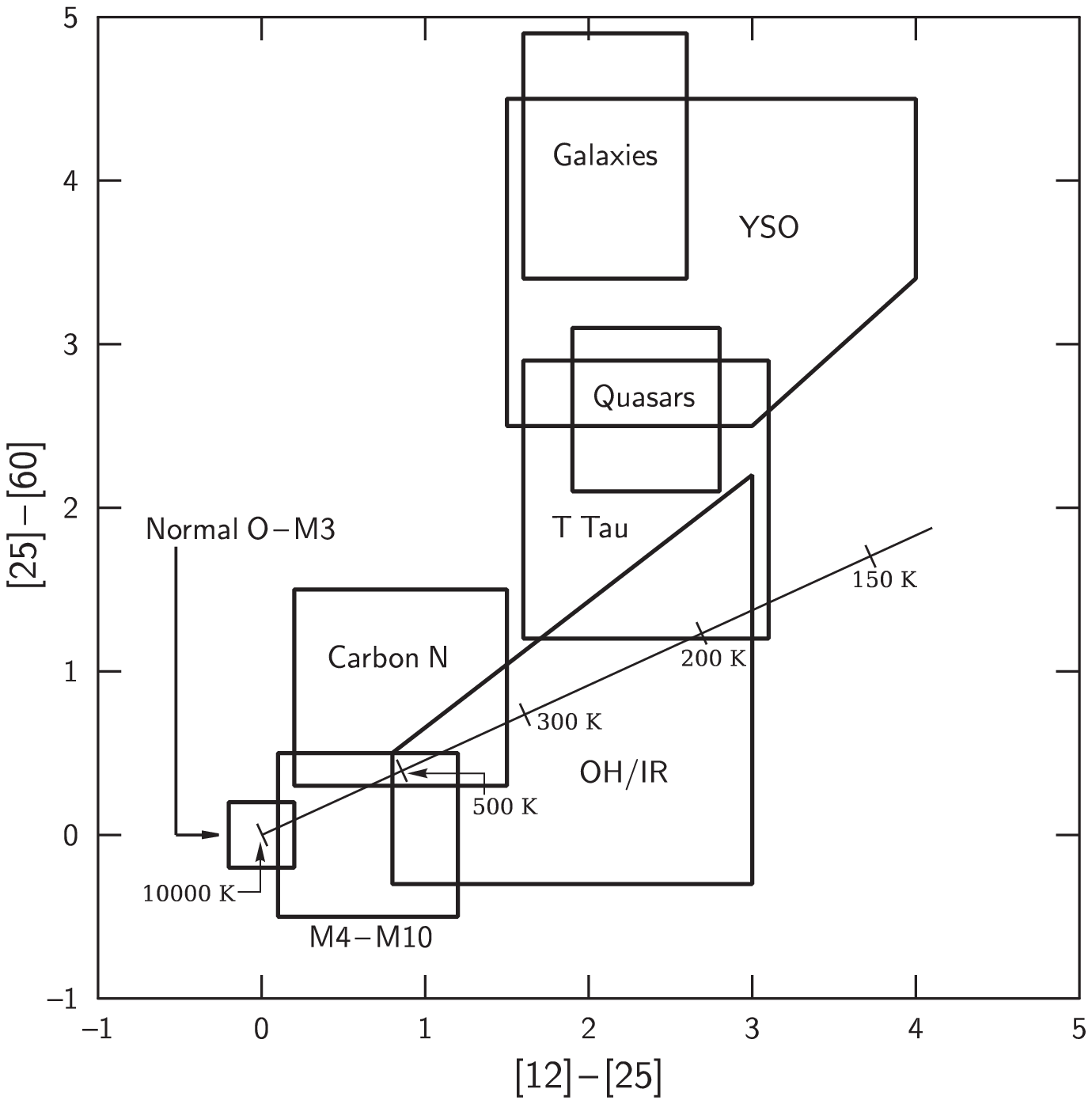,width=115mm,angle=0,clip=true}}
\vspace{0mm}
\captionb{2}{The diagram [12]--[25] vs.  [25]--[60] with the `occupation
zones' of various Galactic and extragalactic objects.  The black-body
line with temperature marks in Kelvins is also shown.}
\end{figure}

In this diagram the `occupation zones' for a broad range of known
stellar and non-stellar sources were shown by Walker \& Cohen (1988) and
Walker et al.  (1989).  More literature sources have been used to define
the zones of M- and N-type stars (Hacking et al. 1985; Kwok et al.
1997), infrared LPVs and OH/IR sources (van der Veen \& Habing 1988;
Kwok et al. 1997; Sevenster 2002; Jim\'enez-Esteban et al. 2006) and
infrared galaxies (Emerson 1987; Magnani et al. 1995).  The zone of T
Tauri stars was taken from Harris et al.  (1988):  [12]--[25] between
1.64 and 3.00, and [25]--[60] between 1.23 and 2.90.  The zone of
infrared YSOs was based on the investigations of Persson \& Campbell
(1987, 1988), Campbell et al.  (1989) and Kenyon et al.  (1990).  The
zones of these types of stars, plotted in Figure~2, were used to
identify and reject the remaining carbon stars, infrared LPVs and other
types of objects unrelated to star forming.  However, even IRAS colors
are insufficient to separate young stellar objects from galaxies,
quasars and the coolest OH/IR sources.  Almost all IRAS galaxies are of
spiral type with active star formation and possess a large amount of
dust.  Their emission in the far infrared is mostly due to thermal
radiation from interstellar dust grains heated by stars.

The stars falling into the IRAS diagram boxes of young stars (T Tauri
and YSO zones) were included in the list of potential young objects.  We
also were able to identify a few young stars having the color index
[12]--[25] of good accuracy but with inaccurate [25]--[60] and vice
versa.  The stars without IRAS observations (or with the IRAS data of
poor quality) were also included in the list:  their belonging to YSOs
was based only on the $J$--$H$ vs.  $H$--$K_s$ diagram and the
concentration to molecular/dust clouds.  The final list of 142 suspected
YSOs is presented in Table 1. Its explanation is given in the next
section.

\landscape
\small
\noindent
\tabcolsep=8pt
\begin{longtable}{rcrrrrrrccll}
\multicolumn{12}{c}{\parbox[c]{165mm}{\baselineskip=10pt
{\bf\ \ Table 1.}{\norm\ Suspected YSOs in the area with $\ell$,\,$b$ =
132--158\degr,
$\pm$\,12\degr\ and $H$--$K_s$\,$\geq$\,1.00.  Abbreviations:
`Gou' means the Gould Belt layer, `Cam' means the Cam OB1 layer and
`Per' means the Perseus arm.\lstrut}}}\\
\firsthline
SL  &  $\ell$ & $b$~~ & $F$~~ & $J$~~ & $H$~~ & $K_s$~ & $J$--$H$ & $H$--$K_s$ & $Q_{JHK}$ & Cloud &  Associated\hstrut \\
    &  deg     &   deg~    &  mag   & mag   &  mag  &  mag  &  mag  &  mag   &   mag            & layer &  objects and notes  \\
\firsthline
\noalign{\vskip1mm}
\endfirsthead
\multicolumn{12}{l}{{\normbf\ \ Table 1.}{\norm\ Continued}}\\
\firsthline
\noalign{\vskip1mm}
SL  &  $\ell$ & $b$~~ & $F$~~ & $J$~~ & $H$~~ & $K_s$~ & $J$--$H$ & $H$--$K_s$ & $Q_{JHK}$ & Cloud &  Associated\hstrut \\
    &  deg     &   deg~    &  mag   & mag   &  mag  &  mag  &  mag  &  mag   &   mag            & layer &  objects and notes  \\
\firsthline
\noalign{\vskip1mm}
\endhead
\endfoot
 1  &  132.015 &   1.215  &  18.02 & 14.46 & 13.09 & 11.93 &  1.37 &  1.16  &   --0.76  &  --   &                           \\[+.5pt]
 2  &  132.124 &   8.884  &  18.17 & 14.28 & 12.87 & 11.80 &  1.41 &  1.07  &   --0.58  &  Cam  &     T878                  \\[+.5pt]
 3  &  132.241 &   8.931  &  18.37 & 14.94 & 13.08 & 11.52 &  1.86 &  1.55  &   --1.01  &  Cam  &     T878                  \\[+.5pt]
 4  &  132.399 &   9.239  &  18.17 & 13.54 & 11.73 & 10.48 &  1.81 &  1.25  &   --0.51  &  Cam  &     T878                  \\[+.5pt]
 5  &  132.824 &   8.944  &  17.31 & 14.44 & 12.74 & 11.70 &  1.70 &  1.05  &   --0.24  &  Cam  &     T878                  \\[+.5pt]
 6  &  132.956 &   8.515  &  18.60 & 14.69 & 13.01 & 11.47 &  1.69 &  1.53  &   --1.15  &  Cam  &     T878                  \\[+.5pt]
 7  &  133.070 & --0.039  &  20.01 & 15.10 & 13.43 & 12.02 &  1.67 &  1.41  &   --0.94  &  Per  &     T879                  \\[+.5pt]
 8  &  133.281 &   8.813  &  18.55 & 14.48 & 12.52 & 11.21 &  1.95 &  1.32  &   --0.48  &  Cam  &     T878                  \\[+.5pt]
 9  &  133.320 &   0.475  &  16.11 & 12.99 & 11.90 & 10.86 &  1.10 &  1.04  &   --0.82  &  Per  &     T879, [1]             \\[+.5pt]
10  &  133.348 &   0.038  &        & 14.36 & 12.41 & 11.04 &  1.95 &  1.37  &   --0.58  &  Per  &     T879, KR 140 [2]      \\[+.5pt]
11  &  133.411 &   0.446  &  19.48 & 14.49 & 12.64 & 11.28 &  1.85 &  1.36  &   --0.67  &  Per  &     [1]                   \\[+.5pt]
12  &  133.411 &   1.195  &        & 14.70 & 12.91 & 11.56 &  1.79 &  1.35  &   --0.71  &  Per  &     T879, W3              \\[+.5pt]
13  &  133.432 &   1.098  &  18.83 & 14.75 & 13.43 & 12.35 &  1.32 &  1.08  &   --0.69  &  Per  &     T879, W3, [1]         \\[+.5pt]
14  &  133.455 &   8.987  &  19.91 & 14.54 & 12.79 & 11.50 &  1.75 &  1.29  &   --0.64  &  Cam  &     T878                  \\[+.5pt]
15  &  133.474 &   0.998  &  19.50 & 13.92 & 12.12 & 11.10 &  1.80 &  1.02  &   --0.09  &  Per  &     T879, W3, [3]         \\[+.5pt]
16  &  133.548 &   0.091  &        & 15.38 & 13.14 & 11.32 &  2.24 &  1.82  &   --1.12  &  Per  &     T879, KR 140 [2]      \\[+.5pt]
17  &  133.572 &   1.087  &        & 15.13 & 13.29 & 12.17 &  1.84 &  1.12  &   --0.23  &  Per  &     T879, W3              \\[+.5pt]
18  &  133.604 &   1.240  &  17.89 & 14.05 & 12.67 & 11.60 &  1.39 &  1.07  &   --0.59  &  Per  &     T879, W3, IC\,1795    \\[+.5pt]
19  &  133.679 &   0.925  &  18.29 & 14.80 & 13.62 & 12.51 &  1.17 &  1.11  &   --0.88  &  Per  &     T879, W3, [3]         \\[+.5pt]
20  &  133.683 &   1.095  &        & 14.98 & 12.67 & 10.91 &  2.31 &  1.75  &   --0.93  &  Per  &     T879, W3              \\[+.5pt]
21  &  133.688 &   0.490  &  15.14 & 12.56 & 11.15 &  9.97 &  1.42 &  1.18  &   --0.76  &  Per  &     T879, [1]             \\[+.5pt]
22  &  133.695 &   1.194  &        & 14.37 & 12.61 & 11.59 &  1.76 &  1.02  &   --0.12  &  Per  &     T879, W3, IC\,1795      \\[+.5pt]
23  &  133.702 &   1.240  &        & 12.86 & 11.08 &  9.62 &  1.78 &  1.46  &   --0.91  &  Per  &     T879, W3, IC\,1795, [3] \\[+.5pt]
24  &  133.705 &   1.199  &        & 13.96 & 11.78 & 10.54 &  2.18 &  1.24  &   --0.10  &  Per  &     T879, W3, IC\,1795, [3] \\[+.5pt]
25  &  133.710 &   1.320  &  19.29 & 14.24 & 12.76 & 11.63 &  1.47 &  1.14  &   --0.63  &  Per  &     T879, W3, IC\,1795      \\[+.5pt]
26  &  133.715 &   1.175  &        & 15.03 & 13.10 & 11.99 &  1.94 &  1.11  &   --0.11  &  Per  &     T879, W3, IC\,1795      \\[+.5pt]
27  &  133.720 &   1.223  &        & 12.06 & 10.04 &  8.84 &  2.02 &  1.19  &   --0.18  &  Per  &     T879, W3, IC\,1795, [3] \\[+.5pt]
28  &  133.723 &   1.259  &  20.02 & 15.20 & 13.63 & 12.60 &  1.57 &  1.03  &   --0.34  &  Per  &     T879, W3, IC\,1795      \\[+.5pt]
29  &  133.739 &   1.228  &        & 14.14 & 11.63 &  9.90 &  2.51 &  1.73  &   --0.68  &  Per  &     T879, W3, IC\,1795      \\[+.5pt]
30  &  133.749 &   1.078  &        & 15.10 & 13.35 & 12.24 &  1.75 &  1.11  &   --0.30  &  Per  &     T879, W3                \\[+.5pt]
31  &  133.766 &   1.320  &        & 14.55 & 12.79 & 11.35 &  1.76 &  1.44  &   --0.90  &  Per  &     T879, W3, IC\,1795, [3] \\[+.5pt]
32  &  133.802 &   1.199  &        & 15.50 & 13.88 & 12.81 &  1.62 &  1.07  &   --0.35  &  Per  &     T879, W3, IC\,1795, [1] \\[+.5pt]
33  &  133.805 &   1.161  &        & 15.47 & 13.92 & 12.84 &  1.55 &  1.08  &   --0.45  &  Per  &     T879, W3, IC\,1795      \\[+.5pt]
34  &  133.809 &   1.217  &        & 11.12 &  8.31 &  6.73 &  2.81 &  1.59  &   --0.13  &  Per  &     T879, W3, IC\,1795      \\[+.5pt]
35  &  133.811 &   1.362  &        & 14.27 & 12.70 & 11.60 &  1.56 &  1.10  &   --0.47  &  Per  &     T879, W3, IC\,1795, [1] \\[+.5pt]
36  &  133.893 &   1.057  &        & 15.21 & 12.94 & 11.60 &  2.28 &  1.33  &   --0.19  &  Per  &     T879, W3, IC\,1795      \\[+.5pt]
37  &  133.919 &   1.204  &  13.31 &  9.70 &  8.60 &  7.58 &  1.11 &  1.02  &   --0.77  &  Per  &     T879, W3, IC\,1795      \\[+.5pt]
38  &  133.994 &   0.520  &        & 15.28 & 13.11 & 11.78 &  2.16 &  1.33  &   --0.30  &  Per  &     T879, [1], [4]          \\[+.5pt]
39  &  134.049 &   0.698  &  19.61 & 14.23 & 12.61 & 11.59 &  1.62 &  1.02  &   --0.26  &  Per  &     T879, LBN 134.05+00.75  \\[+.5pt]
40  &  134.055 &   0.832  &        & 12.94 & 11.51 & 10.49 &  1.43 &  1.02  &   --0.47  &  Per  &     W4, [3]                 \\[+.5pt]
41  &  134.081 &   0.144  &  18.43 & 14.44 & 12.96 & 11.87 &  1.48 &  1.10  &   --0.55  &  Per  &     [1]                     \\[+.5pt]
42  &  134.086 &   0.794  &  20.12 & 15.00 & 13.48 & 12.45 &  1.52 &  1.03  &   --0.40  &  Per  &     W4                      \\[+.5pt]
43  &  134.220 &   0.780  &        & 14.27 & 12.57 & 11.55 &  1.70 &  1.02  &   --0.18  &  Per  &     W4, [1,3]               \\[+.5pt]
44  &  134.234 &   0.751  &        & 13.68 & 11.42 & 10.00 &  2.26 &  1.42  &   --0.37  &  Per  &     W4, [1,6]               \\[+.5pt]
45  &  134.258 & --1.865  &  16.35 & 13.77 & 12.46 & 11.32 &  1.31 &  1.13  &   --0.79  &  Per  &     [6]                     \\[+.5pt]
46  &  134.849 & --1.467  &  17.77 & 14.82 & 13.53 & 12.41 &  1.28 &  1.13  &   --0.80  &  Cam  &                             \\[+.5pt]
47  &  135.053 &   1.663  &  16.68 & 14.75 & 13.50 & 12.31 &  1.26 &  1.18  &   --0.93  &  Per  &     W4                      \\[+.5pt]
48  &  135.517 &   0.255  &        & 12.99 & 11.44 & 10.25 &  1.55 &  1.19  &   --0.66  &  Per  &     W4, [1,6]               \\[+.5pt]
49  &  136.447 &   2.469  &  16.37 & 13.41 & 11.67 & 10.26 &  1.74 &  1.41  &   --0.87  &  Per  &     [1]                     \\[+.5pt]
50  &  136.472 &   1.252  &  19.29 & 14.63 & 12.24 & 10.39 &  2.40 &  1.84  &   --1.02  &  Per  &     T879-B5, W5, [1,5]      \\[+.5pt]
51  &  136.519 &   0.850  &  18.73 & 14.74 & 13.18 & 12.12 &  1.55 &  1.06  &   --0.41  &  Per  &     T879-P5, W5             \\[+.5pt]
52  &  136.673 &   1.207  &        & 14.21 & 10.66 &  8.38 &  3.54 &  2.29  &   --0.69  &  Per  &     T879-P5, W5, [4,5,6]    \\[+.5pt]
53  &  136.796 &   1.083  &  18.41 & 14.67 & 13.34 & 12.22 &  1.34 &  1.11  &   --0.71  &  Per  &     T879-P5, W5             \\[+.5pt]
54  &  136.846 &   1.150  &  17.20 & 11.13 &  8.95 &  7.24 &  2.18 &  1.71  &   --0.98  &  Per  &     T879-P5, W5, [1,4]      \\[+.5pt]
55  &  136.849 &   1.106  &  15.75 & 13.16 & 12.11 & 11.08 &  1.05 &  1.03  &   --0.86  &  Per  &     T879-P5, W5             \\[+.5pt]
56  &  136.917 &   1.085  &        & 12.75 &  9.19 &  6.55 &  3.56 &  2.63  &   --1.31  &  Per  &     W5, [1,4,5,6]           \\[+.5pt]
57  &  137.025 &   1.034  &  19.85 & 14.51 & 13.06 & 11.99 &  1.46 &  1.07  &   --0.52  &  Per  &     W5                      \\[+.5pt]
58\rlap{*} &  137.114 &   3.120  &  17.82 & 14.39 & 12.70 & 11.56 &  1.69 &  1.13  &   --0.41  &  Per  &                             \\[+.5pt]
59  &  137.364 &   0.636  &  17.57 & 14.61 & 13.14 & 12.10 &  1.47 &  1.05  &   --0.47  &  Per  &     W5                      \\[+.5pt]
60  &  137.393 &   0.610  &  16.97 & 13.00 & 11.52 & 10.45 &  1.48 &  1.08  &   --0.51  &  Per  &     W5                      \\[+.5pt]
61  &  137.506 &   1.391  &        & 14.38 & 11.91 & 10.23 &  2.47 &  1.68  &   --0.64  &  Per  &     W5                      \\[+.5pt]
62  &  137.657 &   1.667  &  19.07 & 14.07 & 12.81 & 11.78 &  1.26 &  1.03  &   --0.65  &  Per  &     W5, [1]                 \\[+.5pt]
63\rlap{*} &  137.749 &   1.492  &  16.41 & 12.49 & 11.09 & 10.03 &  1.40 &  1.06  &   --0.57  &  Per  &     W5, [1], BDSB\,58 [7]   \\[+.5pt]
64  &  137.791 &   1.457  &  20.04 & 13.80 & 12.04 & 10.84 &  1.76 &  1.20  &   --0.46  &  Per  &     W5, BDSB\,58 [7]        \\[+.5pt]
65  &  137.920 &   1.924  &  18.63 & 14.74 & 12.94 & 11.59 &  1.79 &  1.35  &   --0.71  &  Per  &     W5, [1]                 \\[+.5pt]
66  &  138.294 &   1.559  &        & 15.40 & 12.92 & 11.38 &  2.48 &  1.54  &   --0.37  &  Per  &     T912, W5, [4,6]         \\[+.5pt]
67  &  138.315 & --0.030  &  16.55 & 14.43 & 12.91 & 11.65 &  1.51 &  1.27  &   --0.83  &  Per  &                             \\[+.5pt]
68\rlap{*} &  138.405 &   3.784  &        & 14.02 & 13.11 & 12.00 &  0.91 &  1.11  &   --1.15  &  Cam  &     [1]       \\[+.5pt]
69  &  138.461 &   1.622  &        & 13.75 & 11.84 & 10.40 &  1.91 &  1.44  &   --0.76  &  Per  &     T912, Sh\,2-201         \\[+.5pt]
70  &  138.488 &   1.641  &        & 14.02 & 11.96 & 10.60 &  2.06 &  1.36  &   --0.46  &  Per  &     T912, W5                \\[+.5pt]
71  &  138.490 &   1.667  &        & 14.23 & 13.05 & 11.85 &  1.18 &  1.20  &   --1.04  &  Per  &     T912, W5                \\[+.5pt]
72  &  138.495 &   1.639  &        & 14.30 & 11.69 & 10.24 &  2.61 &  1.46  &   --0.08  &  Per  &     T912, W5, [6]           \\[+.5pt]
73  &  138.524 &   1.660  &  20.00 & 15.39 & 13.30 & 11.82 &  2.10 &  1.48  &   --0.64  &  Per  &     T912, W5                \\[+.5pt]
74  &  138.533 &   2.059  &  18.93 & 14.53 & 12.84 & 11.74 &  1.69 &  1.10  &   --0.35  &  Per  &                             \\[+.5pt]
75  &  138.550 &   1.580  &  14.18 & 11.67 & 10.06 &  8.83 &  1.60 &  1.23  &   --0.68  &  Per  &     T912, W5                \\[+.5pt]
76  &  138.891 &   5.403  &  19.67 & 15.03 & 13.37 & 12.33 &  1.66 &  1.04  &   --0.26  &  Cam  &     T910                    \\[+.5pt]
77  &  139.320 & --3.306  &  16.74 & 14.97 & 13.40 & 12.11 &  1.58 &  1.28  &   --0.79  &  Per  &                             \\[+.5pt]
78  &  139.369 & --3.283  &  13.10 & 11.15 & 10.08 &  9.02 &  1.07 &  1.06  &   --0.88  &  Cam  &                             \\[+.5pt]
79\rlap{*} &  139.788 &   1.269  &  15.12 & 12.82 & 11.75 & 10.65 &  1.07 &  1.10  &   --0.97  &  Per  &     LBN\,140.07+1.64, [1,5] \\[+.5pt]
80  &  139.911 &   0.196  &        & 12.76 & 10.88 &  9.52 &  1.88 &  1.36  &   --0.64  &  Per  &     LBN\,140.07+1.64, [5,6] \\[+.5pt]
81  &  140.032 &   2.051  &  19.92 & 13.94 & 12.12 & 10.94 &  1.82 &  1.18  &   --0.38  &  Cam  &     Sh\,2-202, LBN\,139.57+2.70, [5] \\[+.5pt]
82\rlap{*} &  140.160 &   2.268  &   9.31 &  9.64 &  8.32 &  7.27 &  1.32 &  1.05  &   --0.62  &  Cam  &     Sh\,2-202, LBN\,139.57+2.70 [5]  \\[+.5pt]
83  &  140.592 &   1.086  &  19.06 & 13.47 & 11.25 &  9.70 &  2.22 &  1.55  &   --0.65  &  Per  &     LBN\,140.07+1.64, [1,5] \\[+.5pt]
84  &  140.600 &   1.122  &  17.21 & 14.77 & 13.47 & 12.38 &  1.30 &  1.09  &   --0.72  &  Per  &     LBN\,140.07+1.64, [5]   \\[+.5pt]
85\rlap{*} &  140.749 &   4.172  &  14.34 & 12.96 & 11.74 & 10.70 &  1.22 &  1.05  &   --0.72  &  Cam  &                             \\[+.5pt]
86  &  140.768 &   0.199  &  18.91 & 12.63 & 10.78 &  9.64 &  1.84 &  1.14  &   --0.28  &  Per  &     LBN\,140.07+1.64, [1,5] \\[+.5pt]
87\rlap{*} &  140.914 & --1.164  &        & 12.88 & 11.53 & 10.53 &  1.35 &  1.00  &   --0.51  &  Per  &     LBN\,140.77-1.42, [1,5] \\[+.5pt]
88  &  141.068 & --1.582  &  17.42 & 13.78 & 12.28 & 11.14 &  1.50 &  1.14  &   --0.60  &  Per  &     LBN\,140.77-1.42, [1,5] \\[+.5pt]
89  &  141.926 &   1.738  &  19.16 & 13.55 & 11.91 & 10.90 &  1.64 &  1.00  &   --0.21  &  Cam  &     T942                    \\[+.5pt]
90  &  141.972 &   0.581  &        & 14.82 & 13.03 & 11.88 &  1.79 &  1.16  &   --0.34  &  Cam  &     T942                    \\[+.5pt]
91  &  141.974 &   1.708  &        & 15.30 & 13.65 & 12.58 &  1.65 &  1.08  &   --0.34  &  Cam  &     T942                    \\[+.5pt]
92  &  141.979 &   1.691  &        & 15.39 & 13.65 & 12.50 &  1.74 &  1.15  &   --0.39  &  Cam  &     T942                    \\[+.5pt]
93  &  141.982 &   1.760  &        & 15.06 & 13.02 & 11.86 &  2.04 &  1.17  &   --0.12  &  Cam  &     T942                    \\[+.5pt]
94  &  141.986 &   1.827  &  18.24 & 13.39 & 11.61 & 10.28 &  1.78 &  1.33  &   --0.69  &  Cam  &     T942                    \\[+.5pt]
95\rlap{*} &  142.000 &   1.820  &  17.67 & 10.95 &  8.08 &  5.72 &  2.87 &  2.36  &   --1.51  &  Cam  &     T942, GL\,490           \\[+.5pt]
96  &  142.013 &   1.697  &        & 14.94 & 12.99 & 11.76 &  1.96 &  1.23  &   --0.31  &  Cam  &     T942                    \\[+.5pt]
97  &  142.040 &   1.850  &        & 14.78 & 13.02 & 12.00 &  1.77 &  1.01  &   --0.11  &  Cam  &     T942                    \\[+.5pt]
98  &  142.343 &   1.354  &  18.01 & 13.20 & 11.88 & 10.77 &  1.32 &  1.10  &   --0.72  &  Cam  &     T942                    \\[+.5pt]
99\rlap{*} &  142.688 &   1.817  &  15.09 & 13.50 & 12.11 & 11.11 &  1.39 &  1.00  &   --0.47  &  Cam  &     T942                    \\[+.5pt]
100 &  142.772 &   1.476  &  17.79 & 13.72 & 12.06 & 10.91 &  1.65 &  1.15  &   --0.48  &  Cam  &     T942                    \\[+.5pt]
101 &  143.152 &   0.479  &  14.90 & 11.71 & 10.77 &  9.77 &  0.94 &  1.00  &   --0.90  &  Cam  &     T942                    \\[+.5pt]
102 &  143.273 &   1.226  &  18.80 & 13.85 & 12.36 & 11.26 &  1.49 &  1.10  &   --0.53  &  Cam  &     T942                    \\[+.5pt]
103 &  143.324 &   1.762  &        & 14.35 & 12.94 & 11.93 &  1.41 &  1.01  &   --0.46  &  Cam  &     T942                    \\[+.5pt]
104 &  143.511 & --1.646  &  19.14 & 14.70 & 13.27 & 12.12 &  1.42 &  1.15  &   --0.70  &  Per  &     Sh\,2-203, BDSB\,59 [7] \\[+.5pt]
105 &  143.638 & --1.589  &  18.20 & 15.18 & 13.98 & 12.97 &  1.20 &  1.00  &   --0.66  &  Per  &     Sh\,2-203, BDSB\,59 [7] \\[+.5pt]
106 &  143.737 & --1.715  &  15.62 & 13.39 & 11.75 & 10.56 &  1.64 &  1.19  &   --0.56  &  Per  &     Sh\,2-203, BDSB\,59 [7] \\[+.5pt]
107 &  143.803 &   0.767  &  20.46 & 15.21 & 13.70 & 12.59 &  1.51 &  1.11  &   --0.55  &  --   &                             \\[+.5pt]
108 &  143.845 & --1.573  &  17.64 & 14.20 & 12.52 & 11.19 &  1.68 &  1.33  &   --0.78  &  Per  &     Sh\,2-203, BDSB\,59 [7] \\[+.5pt]
109\rlap{*} &  144.668 & --0.713  &  17.12 & 13.16 & 11.57 & 10.25 &  1.59 &  1.32  &   --0.84  &  Per  &                             \\[+.5pt]
110 &  144.784 & --1.042  &  18.46 & 14.63 & 13.40 & 12.22 &  1.23 &  1.18  &   --0.95  &  Per  &                             \\[+.5pt]
111 &  145.306 & --0.732  &  18.59 & 13.75 & 12.04 & 10.79 &  1.71 &  1.26  &   --0.61  &  Per  &                             \\[+.5pt]
112 &  146.199 & --1.365  &  15.45 & 12.54 & 11.60 & 10.55 &  0.94 &  1.05  &   --1.00  &  Cam  &                             \\[+.5pt]
113 &  146.839 &   0.109  &  17.14 & 13.69 & 12.53 & 11.48 &  1.16 &  1.05  &   --0.79  &  --   &                             \\[+.5pt]
114 &  147.882 & --0.541  &  17.50 & 13.96 & 12.94 & 11.91 &  1.02 &  1.03  &   --0.89  &  Per  &                             \\[+.5pt]
115 &  148.081 &   0.215  &  17.35 & 13.92 & 12.48 & 10.82 &  1.44 &  1.66  &   --1.63  &  Per  &     FSR\,655 [8]            \\[+.5pt]
116 &  148.105 &   0.139  &  17.44 & 12.55 & 10.92 &  9.65 &  1.63 &  1.27  &   --0.71  &  Per  &     FSR\,655 [8]            \\[+.5pt]
117 &  148.128 &   0.231  &  17.85 & 14.51 & 13.32 & 12.29 &  1.19 &  1.04  &   --0.73  &  Per  &     FSR\,655 [8]            \\[+.5pt]
118 &  148.613 &   2.413  &        & 14.75 & 12.46 & 10.80 &  2.29 &  1.65  &   --0.76  &  Gou  &                             \\[+.5pt]
119 &  148.683 &   1.562  &  17.93 & 14.95 & 13.82 & 12.79 &  1.13 &  1.04  &   --0.78  &  Gou  &                             \\[+.5pt]
120 &  148.856 &   2.021  &        & 15.23 & 13.35 & 11.90 &  1.88 &  1.44  &   --0.78  &  Gou  &                             \\[+.5pt]
121 &  150.340 &   2.914  &  17.51 & 13.05 & 11.63 & 10.53 &  1.42 &  1.10  &   --0.62  &  Gou  &                             \\[+.5pt]
122 &  150.525 & --0.935  &  18.00 & 14.66 & 13.12 & 12.04 &  1.54 &  1.08  &   --0.45  &  Per  &     Sh\,2-206, BDSB\,61 [7] \\[+.5pt]
123 &  150.593 & --0.844  &        & 11.02 &  8.53 &  7.16 &  2.49 &  1.36  &   --0.04  &  Per  &     Sh\,2-206, BDSB\,61 [7] \\[+.5pt]
124\rlap{*} &  150.687 & --0.689  &  15.24 & 11.70 & 10.32 &  9.08 &  1.38 &  1.24  &   --0.92  &  Per  &     Sh\,2-206, BDSB\,63 [7] \\[+.5pt]
125 &  151.229 &   1.026  &  18.04 & 14.26 & 12.55 & 11.43 &  1.72 &  1.12  &   --0.35  &  Per  &                             \\[+.5pt]
126 &  151.371 &   1.879  &  16.01 & 14.54 & 13.28 & 12.24 &  1.26 &  1.04  &   --0.67  &  --   &                             \\[+.5pt]
127 &  151.392 &   1.287  &  16.47 & 14.26 & 12.88 & 11.77 &  1.37 &  1.12  &   --0.69  &  Per  &                             \\[+.5pt]
128 &  151.439 & --0.493  &        & 14.87 & 13.21 & 11.94 &  1.66 &  1.27  &   --0.70  &  Per  &     Sh\,2-209, BDSB\,65 [7] \\[+.5pt]
129 &  151.609 & --0.222  &        & 15.17 & 13.71 & 12.56 &  1.46 &  1.14  &   --0.65  &  Per  &     Sh\,2-209, BDSB\,65 [7] \\[+.5pt]
130\rlap{*} &  151.612 & --0.458  &  15.83 & 10.92 &  8.90 &  7.12 &  2.02 &  1.79  &   --1.28  &  Per  &     Sh\,2-209, BDSB\,65 [7] \\[+.5pt]
131\rlap{*} &  151.725 & --1.292  &  14.00 & 11.30 &  9.77 &  8.54 &  1.53 &  1.23  &   --0.74  &  Cam  &                             \\[+.5pt]
132 &  151.740 & --0.972  &  17.03 &  9.50 &  7.45 &  6.35 &  2.04 &  1.10  &    ~~0.01  &  Cam  &     T1000                   \\[+.5pt]
133 &  153.448 & --1.122  &  19.67 & 15.29 & 13.51 & 12.42 &  1.78 &  1.09  &   --0.25  &  Per  &                             \\[+.5pt]
134 &  154.306 & --0.180  &  18.46 & 14.42 & 12.85 & 11.76 &  1.57 &  1.09  &   --0.44  &  Cam  &                             \\[+.5pt]
135 &  154.588 &   2.047  &  18.93 & 14.88 & 13.69 & 12.66 &  1.19 &  1.03  &   --0.72  &  Gou  &     T1036                   \\[+.5pt]
136 &  155.432 &   0.635  &  14.72 & 12.84 & 11.72 & 10.67 &  1.12 &  1.05  &   --0.82  &  Gou  &                             \\[+.5pt]
137 &  155.633 & --0.617  &  18.53 & 14.40 & 13.04 & 11.86 &  1.37 &  1.18  &   --0.81  &  Gou  &                             \\[+.5pt]
138 &  156.556 & --1.623  &  17.71 & 14.37 & 13.20 & 12.18 &  1.17 &  1.02  &   --0.71  &  Per  &                             \\[+.5pt]
139 &  156.873 & --2.172  &  18.60 & 14.48 & 12.98 & 11.84 &  1.49 &  1.14  &   --0.62  &  Per: &                             \\[+.5pt]
140 &  156.899 & --2.175  &  15.48 & 13.21 & 12.07 & 11.03 &  1.13 &  1.04  &   --0.80  &  Per: &                             \\[+.5pt]
141\rlap{*} &  157.551 & --4.058  &  18.12 & 15.32 & 13.83 & 12.80 &  1.49 &  1.03  &   --0.41  &  Per  &                             \\[+.5pt]
142 &  157.555 & --8.977  &  16.91 & 11.67 &  9.51 &  7.87 &  2.16 &  1.64  &   --0.88  &  Cam  &     T1054                   \\
\hline
\end{longtable}

\moveright5mm
\vbox{\footnotesize
Notes (for the stars marked by asteriscs):

[1]: CO and IRAS sources, Kerton \& Brunt (2003)

[2]: submm and IRAS sources, Kerton et al. (2001)

[3]: embedded IR sources, Elmegreen (1980)

[4]: infrared clusters, Carpenter et al. (2000)

[5]: objects related to bright nebulae at W5 and Sh 2-202, Karr \&
Martin (2003a,b)

[6]: infrared clusters and groups, Bica et al. (2003a)

[7]: infrared clusters, Bica et al. (2003b), BDSB numbers

[8]: infrared clusters, Froebrich et al. (2007), FSR numbers
\vskip2mm


SL\,63. LW Cas, IRAS 02534+6029, INA variable

SL\,82. CPM\,7, IRAS 03134+5958, YSO, Herbig \& Bell (1988), Campbell et al.  (1989)

SL\,95. GL\,490, IRAS 03236+5836, classical YSO

SL\,109. CPM\,8, IRAS 03293+5500, YSO, Campbell et al.  (1989)

SL\,124. GLMP\,49, IRAS 04010+5118, T Tauri star, Garc\'ia-Lario et al. (1997)

SL\,130. CPM\,12, IRAS 04064+5052, YSO, Campbell et al.  (1989)

SL\,141. GLMP\,61, IRAS 04172+4411, T Tauri star, Garc\'ia-Lario et al. (1997)

SL 58, 68, 79, 85, 87, 99, 124 and 131: binaries in R color (SkyView), the
distance between components $<$\,10\arcsec.
}
\endlandscape

\vbox{
\begin{center}
\vbox{\small\tabcolsep=5pt
\parbox{110mm}{\baselineskip=9pt
{\smallbf\ \ Table 2.}{\small\ IRAS and MSX data for the suspected
YSOs in the investigated area with
$H$--$K_s$\,$\geq$\,1.00, supposed to be the Local arm objects. Fluxes
are given in Janskys.
\lstrut}}
\begin{tabular}{rcrrrrr}
\tablerule
 SL &   IRAS       &  $F$(12) &  $F$(25) &  $F$(60) & $F$(100)    & $F$(8.3)  \\
\tablerule
  3 &   02371+6934 &  $<$0.27 &  0.24    &  $<$0.49 &   $<$12.37  &           \\[-1pt]
  4 &   02402+6947 &  $<$0.31 &  0.25    &  0.42    &   $<$16.71  &           \\[-1pt]
  5 &   02432+6919 &  $<$0.37 &  $<$0.25 &  0.55    &   $<$17.04  &           \\[-1pt]
  8 &   02470+6901 &  $<$0.25 &  0.64    &  1.35    &   7.41      &           \\[-1pt]
 14 &              &          &          &          &             &  0.15     \\[-1pt]
 68 &   03074+6211 &  $<$0.25 &  $<$0.50 &  $<$0.40 &   4.46      &           \\[-1pt]
 78 &              &          &          &          &             &  0.19     \\[-1pt]
 81 &   03116+5951 &   0.39   &  0.54    &  $<$2.77 &   27.90     &  0.15     \\[-1pt]
 82 &   03134+5958 &   1.91   &  2.10    &  9.33    &   37.13     &  1.34     \\[-1pt]
 89 &   03228+5834 &   0.24   &  1.06    &  $<$4.04 &   $<$27.71  &           \\[-1pt]
 93\rlap{*} &   03233+5833 &   0.44   &  2.13    &  23.3    &   40:       &           \\[-1pt]
 95\rlap{*} &   03236+5836 &   90.5  &  290  &  715  &  1156    &  54.73    \\[-1pt]
101 &              &          &          &          &             &  0.11     \\[-1pt]
102 &   03290+5724 &  $<$0.26 &  0.17    &  $<$0.75 &   $<$44.76  &           \\[-1pt]
107 &   03303+5643 &  $<$0.39 &  $<$0.25 &   0.69   &   $<$5.80   &           \\[-1pt]
108 &   03211+5446 &  12.67   &  43.09   &  496.5   &   854.3     &           \\[-1pt]
112 &   03353+5333 &   0.28   &  $<$0.57 &  $<$0.40 &   $<$18.18  &  0.15     \\[-1pt]
113 &   03447+5422 &  $<$0.26 &  0.31    &  0.46    &   $<$12.64  &           \\[-1pt]
118 &   04044+5500 &  $<$0.25 &  0.55    &  2.02    &   $<$11.02  &  0.25     \\[-1pt]
120 &   04038+5433 &   0.37   &  0.48    &  1.19    &   $<$5.37   &           \\[-1pt]
126 &   04156+5244 &  $<$0.35 &  0.40    &  1.49    &   $<$25.57  &  $<$0.08  \\[-1pt]
131 &   04034+5010 &    1.93  &  2.89    &  5.50    &   19.23     &  1.30     \\[-1pt]
132 &              &          &          &          &             &  0.58     \\[-1pt]
134 &   04198+4912 &  $<$0.75 &  $<$0.42 &  0.57    &   $<$3.12   &           \\[-1pt]
135 &   04308+5033 &   0.28   &  0.48    &  2.53    &   $<$14.19  &           \\[-1pt]
137 &   04235+4757 &  $<$0.35 &  0.67    &  2.00    &   $<$5.49   &           \\[-1pt]
142 &   03591+4034 &   1.30   &  $<$0.40 &  $<$0.40 &   $<$7.70   &           \\
\noalign{\vskip0.5mm}
\tablerule
\end{tabular}
}
\end{center}
}
\moveright10mm
\vbox{\footnotesize
Notes:

Stars 93 and 95: IRAS data are from Clark (1991). }

\sectionb{3}{ATTRIBUTION OF YSOs TO DIFFERENT CLOUD LAYERS}

\enlargethispage{4mm}

In Paper I we divided molecular/dust clouds in the area by their CO
radial velocities (from Dame et al. 2001) to the following three layers:
the Gould Belt layer at 150--300 pc from the Sun, the Cam OB1
association layer at 800--900 pc and the Perseus arm at 2--3 kpc (see
the $\ell$,\,$b$ map in Fig.\,6 of Paper I).  The objects from Table 1
were plotted on this map to attribute them to the listed dust layers.
Most of the selected YSOs (68 objects) located within the longitudes
132--140\degr\ and latitudes $\pm$\,2\degr\ depend to the Perseus arm
and coincide with the dust clouds DoH 879 and 912 related to the H\,II
regions W3, W4 and W5 and the Cas OB6 association (the identification of
the Dobashi cloud numbers is given in Paper I, Fig.\,2).  Additional 26
objects are assigned to the Perseus arm at larger longitudes.

Our attribution of YSOs to certain molecular/dust cloud layers was
confirmed by inspection of the LSR radial velocities of the associated
CO clouds presented by Wouterloot \& Brand (1989), Wouterloot et al.
(1993) and Kerton \& Brunt (2003).

Table 1 for each object gives the serial number, Galactic coordinates,
the $F$ magnitudes (close to $R$) taken from the GSC 2.3.2 catalog
available at the CDS, the 2MASS {\it J, H} and $K_s$ magnitudes, color
indices and $Q_{JHK}$ parameters, belonging to the assigned dust/gas
layer (Per, Cam or Gou, see the Table head for explanation) and the
associated dark or bright clouds and clusters.  The Cam OB1 layer
contains 35 objects, 16 of them are concentrated in the cloud DoH 942
(clumps P1, P2 and P3) and seven in the cloud DoH 878 at 133\degr,
+9\degr.  The Gould Belt layer contains only seven objects, all located
between $\ell$\,=\,148\degr\ and 156\degr.  For a few objects we could
not assign a single layer due to overlapping of the clouds.

In Figure 3 the objects from Table 1 are plotted in Galactic
coordinates, together with the Dobashi dark clouds (Dobashi et al.
2005).  Here we see an obvious grouping of the suspected YSOs to the
darkest dust clouds, and this is one of the indications that these
objects may be young.  Some of the YSOs are projected close to the
infrared open clusters identified by Carpenter et al.  (2000), Bica et
al.  (2003a,b) and Froebrich et al.  (2007):  five objects near the Bica
et al. cluster No.\,59, four objects near the clusters Bica et al.
No.\,61 and Froebrich et al.  No.\,665 (in the direction of the H\,II
region Sh 2-206), and three objects near the cluster Bica et al.
No.\,65 (in the direction of Sh 2-209).

All objects in Table 1 were inspected in the red plates of the POSS as
well as in the 2MASS passbands, presented in the SkyView site of NASA
(http://skyview.\\ gsfc.nasa.gov).  Notes on the double or oblong images
are given at the end of Table~1.


\begin{figure}[!th]
\centerline{\psfig{figure=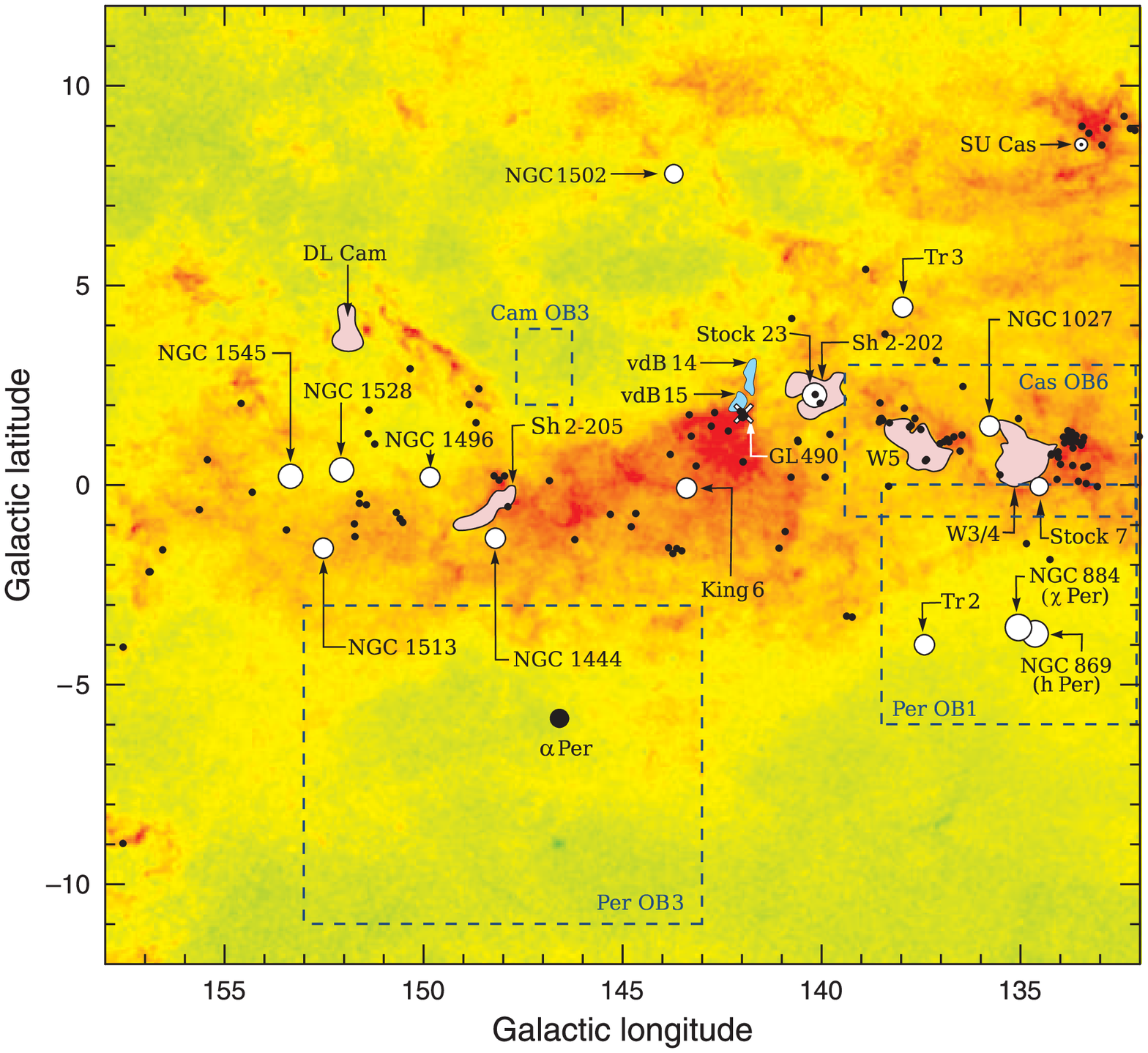,width=124mm,angle=0,clip=true}}
\vspace{0mm}
\captionb{3} { The suspected YSOs from Table 1 (black dots) plotted
together with the dust clouds from Dabashi et al.  (2005), nebulae and
open clusters belonging to the Local arm (plus the h+$\chi$ Per
cluster).  The object GL\,490 is shown as the white cross (and the white
arrow).  The four rectangles are boundaries of the associations Cas OB6
and Per OB1, located in the Perseus arm, Cam OB3 in the Outer arm and
Per OB3 in the foreground of Cam OB1.  The stars of Cam OB1 are
scattered over the whole area, see Paper I.}
\end{figure}

In further analysis we will consider only the Local arm objects.  YSOs
in the Cassiopeia section of the Perseus arm were already investigated
in many papers:  Elmegreen (1980), Megeath et al.  (1996), Deharveng et
al.  (1997), Carpenter et al.  (2000), Ogura et al.  (2002), Karr \&
Martin (2003a,b), Ojha et al.  (2004), Ruch et al.  (2007), Saito et al.
(2007), etc.

In Table 2 we list only 27 YSOs belonging to the Local arm and matched
with the IRAS and/or MSX point sources.  For them we give IRAS numbers
and fluxes in Janskys in the 12, 25, 60 and 100 $\mu$m passbands,
collected from the Simbad database.  For some of the objects the fluxes
in the 8.3 $\mu$m passband of MSX (Egan et al. 2003) are given too.  In
other MSX passbands the fluxes for most objects are of lower accuracy
and have not been used.

The objects of the Local arm listed in Table 1 are plotted in the
$J$--$H$ vs.  $H$--$K_s$ diagram (Figure 4) as red dots.  Other young
objects of the area:  O--B3 stars of Cam OB1 (black dots), irregular
variables (blue dots) and H$\alpha$ emission stars (blue circles), are
also plotted taking their data from Paper I. The star LW Cas was
excluded as it belongs to the Perseus arm.  The sharp edge of the red
dots is the result of the selection criterion ($H$--$K_s$\,$\geq$\,1.0).
In reality, more young objects are expected within $H$--$K_s$ between
0.4 and 1.0 but it is difficult to identify them among thousands of
other stars lying in the same area of the diagram.  The distribution of
the suspected YSOs in the $J$--$H$ vs.  $H$--$K_s$ diagram for
Camelopardalis is quite similar to that in other star-forming regions
(see, e.g., Lada \& Adams 1992; Gomez et al. 1994 and Kenyon \& Hartmann
1995 for pre-main-sequence stars in Taurus).

\newpage

\sectionb{4}{SPECTRAL ENERGY DISTRIBUTIONS}


\begin{figure}[!th]
\centerline{\psfig{figure=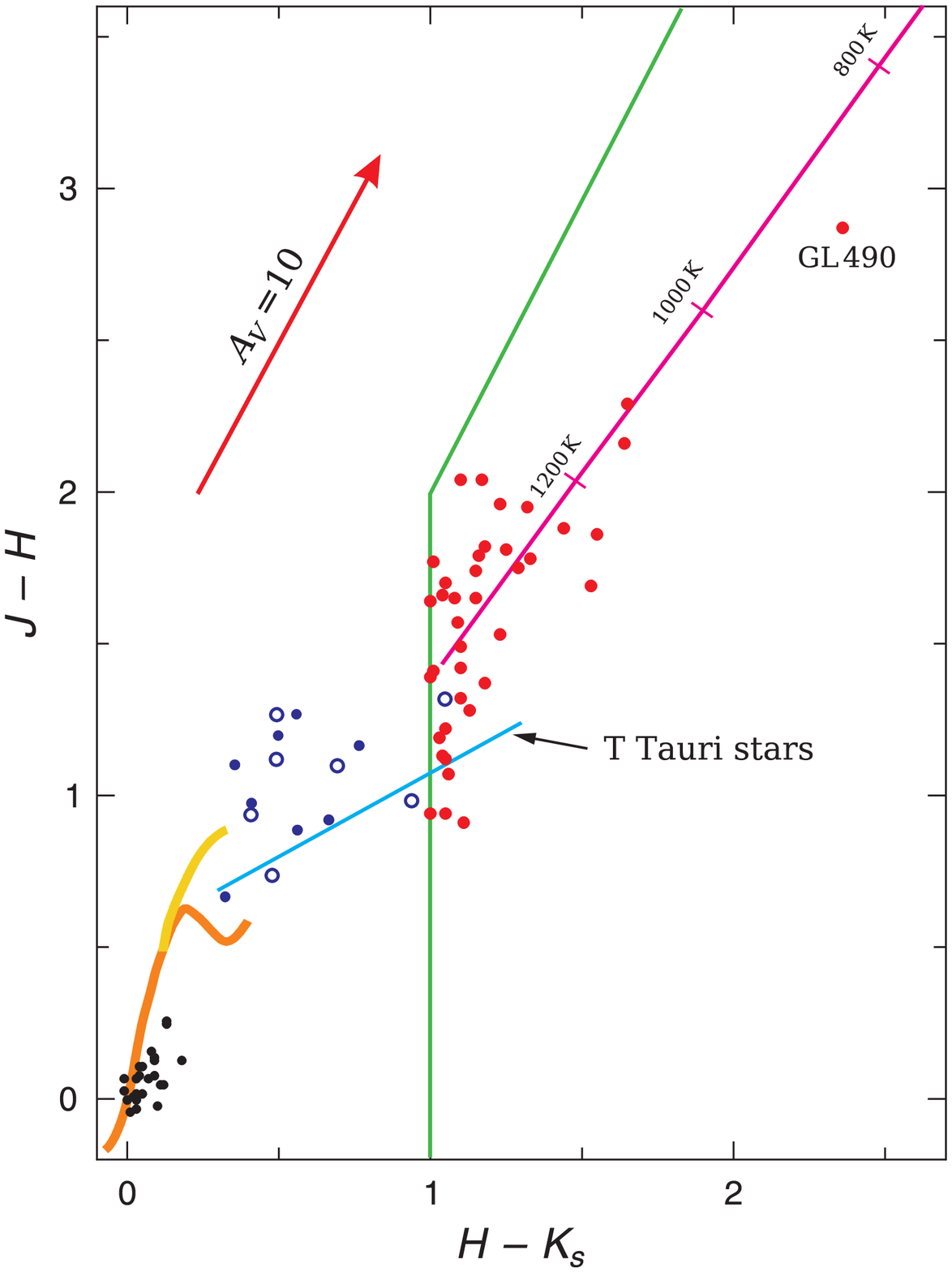,width=95mm,angle=0,clip=true}}
\vspace{.8mm}
\captionb{4}{ The same as in Figure 1, but only with real and suspected
young objects belonging to the Local arm, i.e., to the `Cam' and `Gou'
cloud layers (red dots).  Black dots designate O--B3 stars of the Cam
OB1 association, blue dots designate known irregular variables and blue
open circles the H$\alpha$ emission stars (see Paper I).}
\end{figure}

In the $J$--$H$ vs.  $H$--$K_s$ diagram the YSO sequence contains
pre-main-sequence stars at various stages of evolution and having
various physical parameters.  Lada (1987) has suggested a scheme of
classification for pre-main-sequence stars in three classes, based on
their far-infrared spectral energy distribution (SED) curves in the
$\log\,(\lambda F_{\lambda})$ vs.  $\log\,\lambda$ plane.  Class I
sources exhibit SEDs with a steep rise in the near infrared (up to 10
$\mu$m) and a slower rise up to 100 $\mu$m.  Class II sources exhibit
broad SEDs with a steep rise up to $\sim$\,2 $\mu$m and a shallow
falling-off up to 100 $\mu$m.  Class III objects have SEDs with maxima
at $\sim$\,2 $\mu$m, and their shape is similar to heavily reddened SEDs
of normal stars or black bodies.  Additional Class 0 (zero) was
introduced by Andr\'e et al.  (1993) to designate extremely young submm
continuum objects or protostars.  Evolutionary interpretation of these
classes, given by Adams et al.  (1987), is still valid, but now we have
much more complete observational data and a better theoretical
understanding of early stages of stellar evolution.

Whitney et al.  (2003a,b, 2004) and Robitaille et al.  (2006, 2007) have
calculated a three-dimensional grid of YSO models covering a wide range
of stellar masses, evolutionary stages and viewing angles.  Their Stage
I objects have significant infalling envelopes and possibly disks, Stage
II objects have optically thick disks (and possible remains of tenuous
infalling envelope), and Stage III objects have optically thin disks.
Typical Stage II objects are classical T Tauri-type stars with strong
emission lines (CTTS), and typical Stage III objects are weak-lined T
Tauri stars (WTTS). Since YSOs of Stages II and III have their own
names (CTTS and WTTS), the term YSO in many times is used to designate
only the objects of Stage I of Robitaille et al. (2006) or the objects
of Classes I and 0 of Lada (1987) and Andr\'e et al. (1993). In this
paper we use the term YSO for all pre-main-sequence objects.

The distribution of YSO models in the $J$--$H$ vs.  $H$--$K$ diagram
(Robitaille et al. 2006) shows close resemblance to the diagrams based
on observations of YSOs in star-forming regions, including our Figure 4.
Stage I objects tend to be located at the cool end of the sequence, with
$H$--$K$\,$>$\,1.  GL\,490 in Camelopardalis is one of such objects.
Most of the YSOs with central stars of high temperatures ($>$\,10\,000
K) also have colors redder than those of low-mass objects.  However,
in the presence of different and unknown interstellar and circumstellar
reddening, identification of YSOs in different stages of evolution using
only the $J$--$H$ vs.  $H$--$K_s$ diagram is quite complicated.

SEDs in the infrared are the most reliable tools for separating YSOs
from normal (and, partly, from AGB) stars.  The photospheres of normal
and AGB stars having no dense envelopes of low temperatures exhibit
black-body like curves in the plot $\log \lambda\,F(\lambda)$ vs.  $\log
\lambda$, with sharp maxima at $\sim$\,2 $\mu$m and almost a linear drop
of the flux with increasing wavelength; see Figure 5, where SEDs for
intrinsic and heavily reddened K0\,III, M0\,III and M5\,III stars are
given.  Interstellar reddening shifts the flux maximum towards longer
wavelengths making the long-wave slope of SED curves steeper.  This
means that normal stars even with extremely heavy reddening remain only
 near infrared


\begin{wrapfigure}[30]{r}[0pt]{62mm}
\vskip-3mm
\psfig{figure=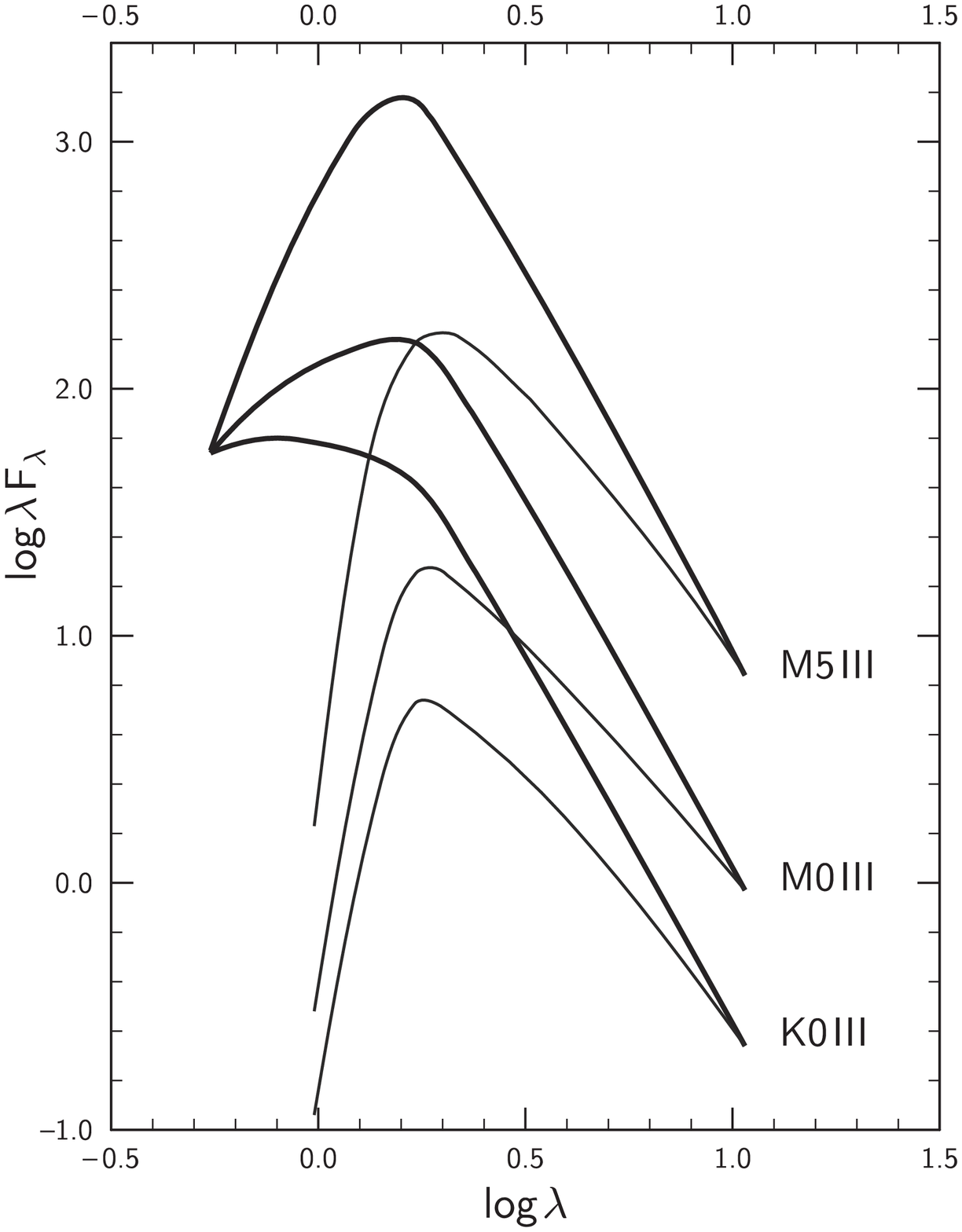,width=61mm,angle=0,clip=true}
\vspace{.8mm}
\vskip-2mm
\captionr{5}{Spectral energy distributions of K0\,III, M0\,III and
M5\,III stars:  thick lines are for the unreddened stars and thin lines
are for the same stars reddened by the interstellar dust up to the
extinction of 15 mag in the $V$ passband.  The SEDs are obtained from
the intrinsic color indices in the {\it VRIJHKLMN} system and normalized
at 550 nm ($\log\,\lambda$ = --0.26).}
\end{wrapfigure}

\noindent sources.  The flat or rising up SED at the wavelengths longer
than 2 $\mu$m argues for the presence of a circumstellar disk or
envelope.

Oxygen-rich AGB stars with thick dust envelopes (infrared miras and
OH/IR sources) also exhibit excess radiation in the far-infrared, but
their SEDs are quite different from YSOs -- they resemble to SEDs of
black bodies of very low temperatures ($<$\,300 K) with the maximum flux
density between 5 and 50 $\mu$m and the narrow 9.8 $\mu$m and 20 $\mu$m
absorption bands of silicates (see Habing 1996).  In thick dust
envelopes of carbon miras, a SiC band at 11.2 $\mu$m is observed.
However, all these bands cannot be discerned in broad-band photometric
data.

To construct SEDs from the available data of IR surveys,
we calculated $\log \lambda F_{\lambda}$ quantities for the Table 2
stars using the following equations:
$$
\log \lambda F_{\lambda} = \log\,(10^{-m/2.5} \times \lambda \times
F_{\lambda}^{m=0})
$$
for the $R$, $J$, $H$ and $K_s$ magnitudes and
$$
\log \lambda F_{\lambda} = \log (3\,\times\,10^{-9} \times \lambda^{-1}
\times F_{\nu}).
$$
for the [8.3], [12], [25], [60] and [100] fluxes.
Here $\lambda$ is in $\mu$m, $F_{\nu}$ in Jy and $F_{\lambda}$ in
erg\,$\times$\,cm$^{-2}$\,$\times$\,s$^{-1}$\,$\times$\,$\mu$m$^{-1}$.
The following values of $F_{\lambda}$ for $m$ = 0 were taken:
1.66\,$\times$\,10$^{-5}$ for $R$, 3.03\,$\times$\,10$^{-6}$ for
$J$, 1.26\,$\times$\,10$^{-6}$ for $H$ and 4.06\,$\times$\,10$^{-7}$
for $K_s$ (Campins et al. 1985; Strai\v zys 1992). For  the $R$, $J$,
$H$ and $K_s$ magnitudes the mean wavelengths 0.71, 1.26, 1.60 and
2.2 $\mu$m were taken.

Although IRAS measurements are available for 23 objects, in
many cases only the upper limit of the flux is given.  Consequently, we
were not able to construct SEDs from 1.26 to 100 $\mu$m containing all
the IRAS points.  For 10 objects the MSX fluxes at 8.3 $\mu$m were
helpful.

Figures 6--8 present SEDs only for 16 objects for which reliable fluxes
are available at least at some of the IRAS or MSX passbands.  For all
the objects in Figures 6 and 7 the ratio of $F_{12}/F_{25}$ is lower
than 1.0 or [12]--[25]\,$>$\,1.56, and this indicates that radiation of
their photospheres is modified by the circumstellar dust.  We have
attempted to differentiate the objects by the form of their SEDs.
Although in the figures we joined the dots by rounded curves, the waves
in most cases are not real, probably they are a consequence of the
scatter of points due to measurement errors or other reasons.  In many
cases YSOs are not dot-like objects, and the measured fluxes depend on
the size of photometer's aperture used.  The IRAS dots at 60 and 100
$\mu$m are affected by strong thermal dust emission since most of the
YSOs are immersed in dense dust clouds.

In Figure 6 the YSOs, most similar to the Class I objects, are
displayed.  Here the most outstanding object is GL\,490 with extremely
steep flux distribution curve in the near infrared, a small dip near 10
$\mu$m and the maximum at 50--60 $\mu$m.  This means that we are
observing a typical Class I object with dominant thermal radiation from
the envelope heated by a source of high temperature and having a small
inclination of its rotational axis (see models in Fig.\,3 of Whitney et
al. 2004).  The curves of other YSOs in Figure 6 are similar to the
model curves of the Taurus SFR stars IRAS 04169+2702 and IRAS 04181+2654
(see Fig.\,7 of Kenyon et al. 1993 and Fig.\,1 of Robitaille et al.
2007).  The model SEDs correspond to rotationally flattened infalling
envelope and a flared disk with small inclination.  In some cases we
should take into account the strong interstellar reddening in front of
the object, which may considerably reduce the radiation shorter than 1.0
$\mu$m, leaving the radiation longer than 10 $\mu$m not affected.  This
can reduce considerably the required opacity in the circumstellar
envelope.

Figure 7 shows the SEDs for six objects supposed to belong to Class II.
They exhibit a moderate decline of the shortward wing and a more or less
flat spectrum longward of 2--3 $\mu$m.  In YSOs of Class II either
direct radiation from the photosphere or its scattered radiation by the
thick disk should be observed.  Thus, their SEDs should extend into the
optical spectrum (see Whitney et al. 2003b; Robitaille et al. 2006,
2007).  In Figure 7 we see a very faint intensity in the optical
spectrum and no maximum at 1--2 $\mu$m predicted by models.  Both
optical spectrum and the maximum in the near infrared probably are cut
down by strong interstellar reddening.

Figure 8 shows the SEDs for three objects, which are quite similar to
photospheric spectra.  These objects may be either normal stars affected
by heavy reddening or post-T Tauri stars having no optically detectable
circumstellar disk.

In Figure 9 we show the SEDs for five objects belonging to the Perseus
arm, whose ~dependence to the YSO category is generally recognized.
These objects are CPM\,8 and 12 ~(Campbell et al. 1989), GLMP 49 and
61 ~(Garc\'ia-Lario et al.


\vbox{
\hbox{\parbox[t]{85mm}{\psfig{figure=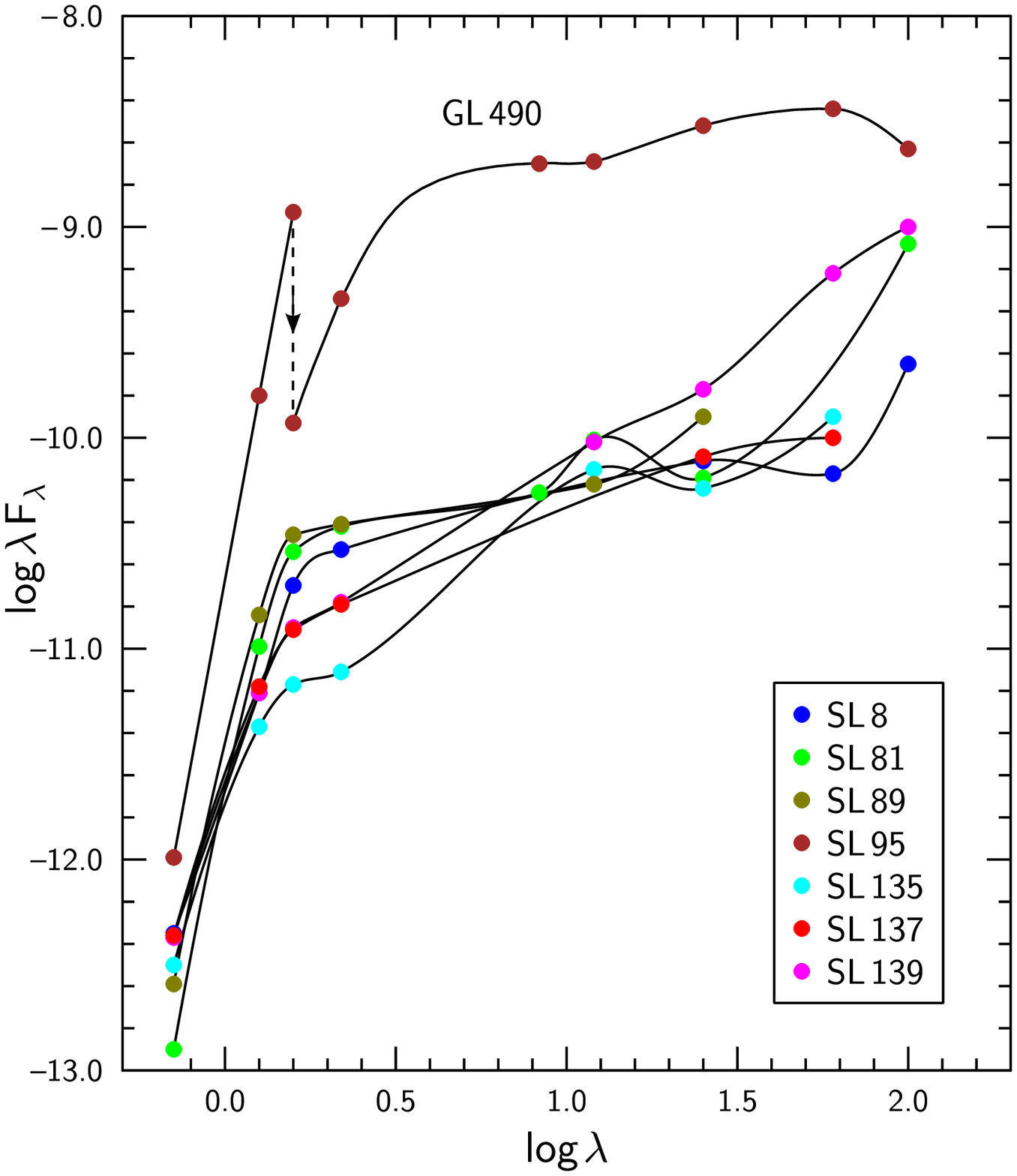,width=80mm,angle=0,clip=true}}
\parbox[t]{37mm}{\vskip-2.5cm\captionr{6}{Spectral energy distributions for seven objects of Table 2
which are most similar to the Class I YSOs.}}}
\vskip5mm
\hbox{\parbox[t]{85mm}{\psfig{figure=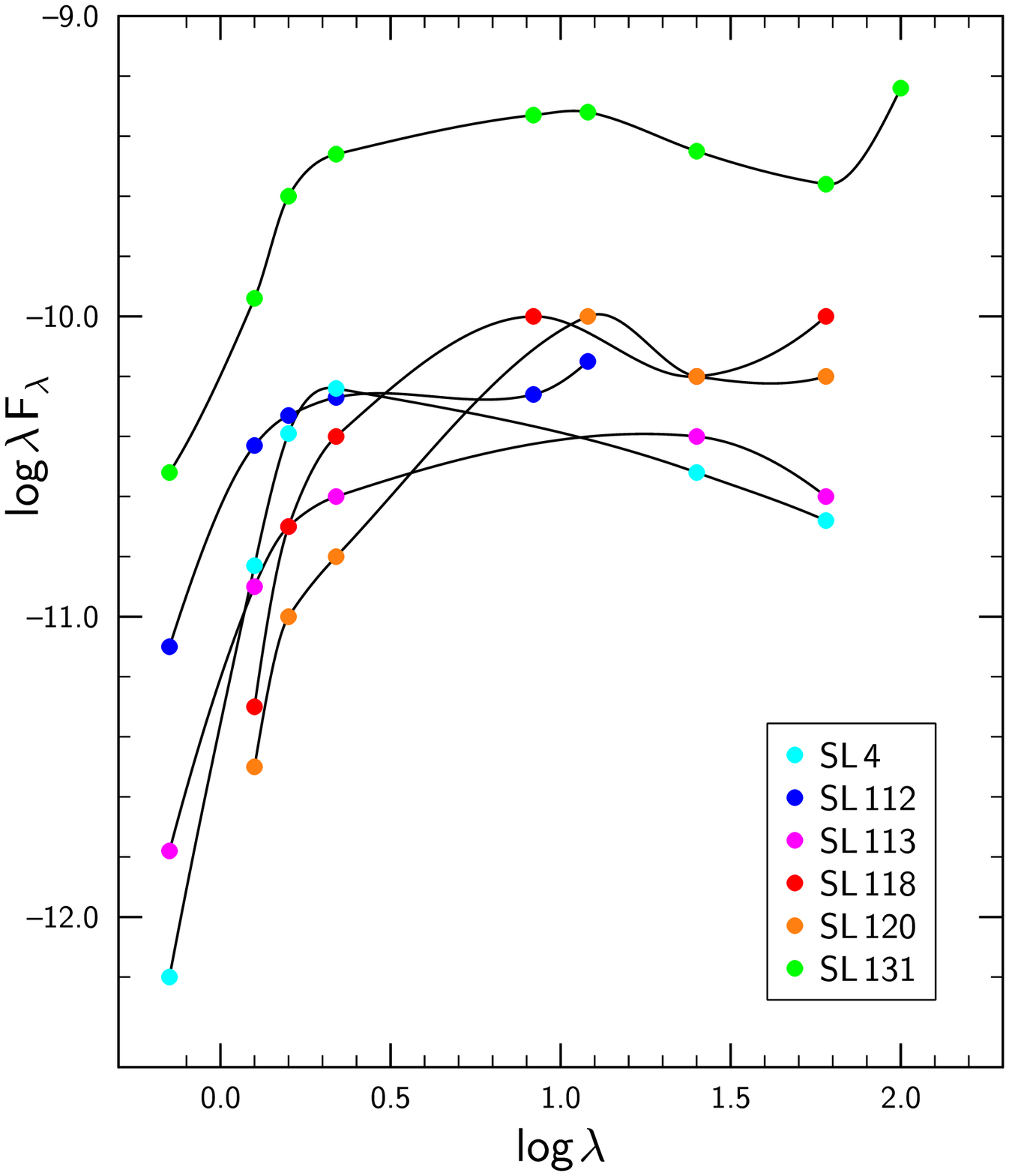,width=80mm,angle=0,clip=true}}
\parbox[t]{37mm}{\vskip-2.5cm\captionr{7}{Spectral energy distributions for six objects of Table 2
which are most similar to the Class II YSOs.}}}}

\vbox{
\hbox{\parbox[t]{85mm}{\psfig{figure=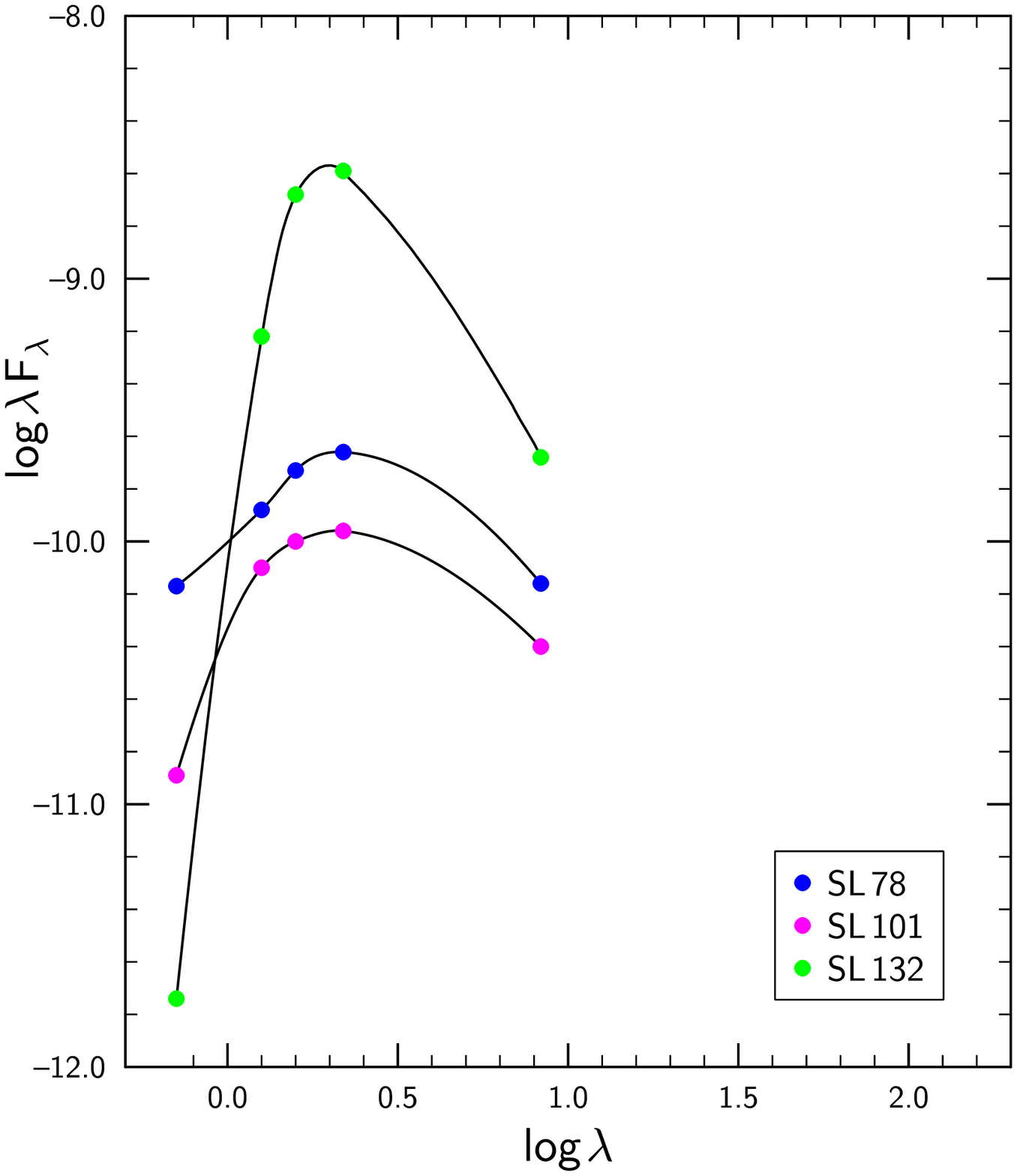,width=80mm,angle=0,clip=true}}
\parbox[t]{37mm}{\vskip-2.5cm\captionr{8}{Spectral energy distributions for three objects of Table 2
which are most similar to the Class III YSOs.}}}
\vskip5mm
\hbox{\parbox[t]{85mm}{\psfig{figure=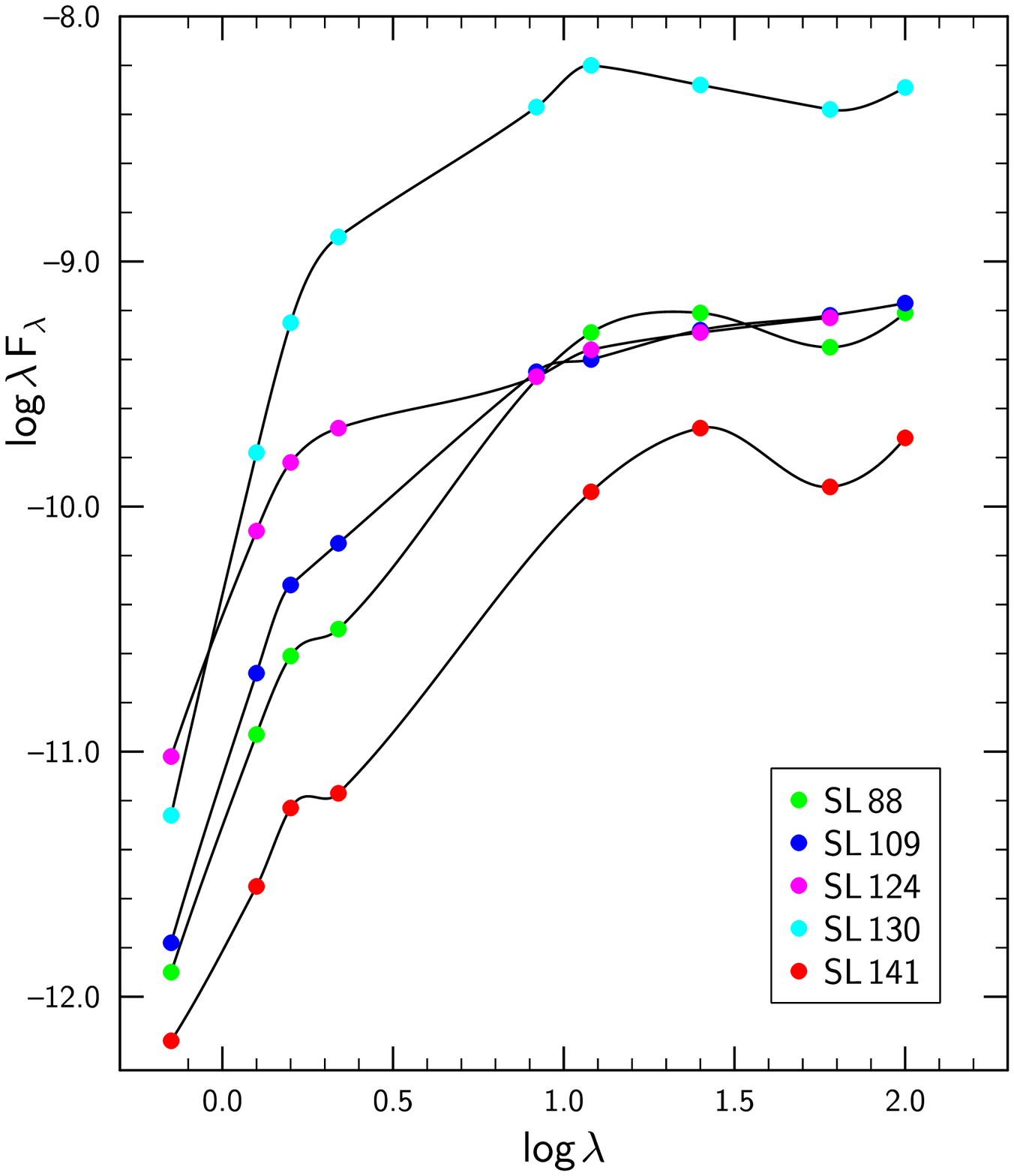,width=80mm,angle=0,clip=true}}
\parbox[t]{37mm}{\vskip-3cm\captionr{9}{Spectral energy distributions for five objects of Table 1
belonging to the Perseus spiral arm. All these objects are confirmed YSOs.}}}}

\noindent 1997), and IRAS 03045+5612 (Kerton \& Brunt 2003).  The SEDs
show that all five objects belong either to Class I or are intermediate
between Classes I and II.  The curve of CPM\,12 at $\lambda >$\,2 $\mu$m
is quite similar that of GL\,490 (Figure 6).

We cannot exclude the possibility that some of the objects described can
be unidentified spiral galaxies (or quasars) with heavy reddening by
dust clouds in our Galaxy.  Many of such galaxies have been already
found behind the Milky Way (see, e.g., Takata et al. 1994; Saurer et al.
1997).  A higher probability is to find some reddened galaxies among the
fainter objects of Table 1, for which no reliable IRAS and MSX data are
available.  Fortunately, the number density of bright galaxies and
quasars is relatively low.  Also, many galaxies are not point sources
and could be recognized in the SkyView images, especially in the optical
passbands.  Probably most (if not all) the SEDs shown in Figures 6--8
belong to young stellar objects.

\sectionb{5}{CONCLUSIONS}

In the Camelopardalis section of the Milky Way, including the nearby
regions of Cassiopeia, Perseus and Auriga, we have identified 142
infrared objects suspected to be in the pre-main-sequence stage of
evolution. The criteria for the attribution to YSOs are their positions
in the $J$--$H$ vs.  $H$--$K_s$ and [12]--[25] vs.  [25]--[60] diagrams,
the similarity of the observed and the model infrared energy
distributions and their concentration in the densest parts of
interstellar molecular and dust clouds.  Approximate distances of the
objects are estimated by radial velocities of the associated CO clouds.
The majority of the identified objects belong to the Perseus arm and
reside in the star-forming region associated with the complex of
emission nebulae W3/W4/W5 and the Cas OB6 association.  Many of these
infrared objects are already known as YSOs from earlier investigations
based on the IRAS, MSX, 2MASS and {\it Spitzer} surveys.

42 YSOs are found to belong to the Local arm:  35 objects are related to
the Cam OB1 association CO layer at a distance of about 1 kpc and seven
to the local CO layer related to the Gould Belt.  About half of the
identified objects of the Cam OB1 layer are concentrated in the dust
cloud DoH 942 (clumps P1, P2 and P3), in the center of the association.
The brightest among them is GL\,490, a well known pre-stellar object of
Class I, which has had numerous investigations in optical, infrared and
radio wavelengths.  Other groups of the detected objects concentrate
either in other dark clouds or in the infrared clusters discovered
recently through 2MASS photometry and infrared imaging surveys.

We do not exclude, however, that some of the objects, suspected to be in
the pre-main-sequence stage, in reality are heavily reddened OB stars
(including Be stars), AGB stars or even distant  galaxies and
quasars since {\it J, H, K} photometry is not sufficient to
differentiate these objects.  Although IRAS and MSX photometry is
helpful in rejecting reddened OB, Be stars and most of the late-type AGB
objects, the dusty spiral galaxies and quasars remain mixed with YSOs.

\thanks {We are thankful to Bo Reipurth, Patricia Whitelock and
\v{Z}eljko Ivezi\'c for consultations and to Stanislava Barta\v si\=
ut\.e and Edmundas Mei\v{s}tas for their help preparing the paper.  The
use of the 2MASS, IRAS, MSX, SkyView and Simbad databases is
acknowledged.}

\References

\refb Adams F. C., Lada C. J., Shu F. H. 1987, ApJ, 312, 788

\refb Andr\'e P., Ward-Thompson D., Barsony M. 1993, ApJ, 406, 122

\refb Beichman C. A., Neugebauer G., Habing H. J., Clegg P. E., Chester
T. J. 1988, {\it IRAS Catalogs and Atlases: Explanatory Supplement},
Washington, DC: GPO, http://irsa.ipac.caltech.edu/applications/Gator/


\refb Bica E., Dutra C. M., Barbuy B. 2003a, A\&A, 397, 177

\refb Bica E., Dutra C. M., Soares J., Barbuy B. 2003b, A\&A, 404, 223



\refb Campins H., Rieke G. H., Lebofsky M. J. 1985, AJ, 90, 896

\refb Campbell B., Persson S. E., Matthews K. 1989, AJ, 98, 643


\refb Carpenter J. M., Heyer M. H., Snell R. L. 2000, ApJS, 130, 381

\refb Clark F. O. 1991, ApJS, 75, 611

\refb Cutri R. M., Skrutskie M. F., Van Dyk S., Beichman C. A. et al.
2003, {\it 2MASS All Sky Catalog of Point Sources}, NASA/IPAC Infrared
Science Archive,\\ http://irsa.ipac.caltech.edu/applications/Gator/

\refb Dame T. M., Hartmann D.,  Thaddeus P. 2001, ApJ, 547, 792

\refb Daugherty S. M., Waters L.\,B.\,F.\,M., Burki G. et al. 1994,
A\&A, 290, 609

\refb Deharveng L., Zavagno A., Cruz-Gonz\'alez I., Salas L., Caplan J.,
Carrasco L. 1997, A\&A, 317, 459

\refb Dobashi K., Uehara H., Kandori R., Sakurai T., Kaiden M.,
Umemoto T., Sato F. 2005, PASJ, 57, S1

\refb Egan M. P., Price S. D., Kraemer K. E. et al. 2003, {\it The
Midcourse Space Experiment Point Source Catalog}, version 2.3,
AFRL-VS-TR-2003-1589; available at CDS: MSX6C Infrared Point Source
Catalog, V/144

\refb Elmegreen D. M. 1980, ApJ, 240, 846

\refb Emerson J. P. 1987, in {\it Star Forming Regions} (IAU Symp.
115), eds.  M. Peimbert \& J. Jugaku, Reidel Publ.  Comp., Dordrecht,
p.\,19

\refb Finlator K., Ivezi\'c \v{Z}., Fan X., Strauss M. A. et al. 2000,
AJ, 120, 2615

\refb Froebrich D., Scholz A., Raftery C. L. 2007, MNRAS, 374, 399

\refb Garc\'ia-Lario P., Manchado A., Pych W., Pottasch S. R. 1997,
A\&AS, 126, 479

\refb Gomez M., Kenyon S. J., Hartmann L. 1994, AJ, 107, 1850

\refb Habing H. J. 1996, A\&ARv, 7, 97

\refb Hacking F., Neugebauer G., Emerson J. et al. 1985, PASP, 97, 616

\refb Harris S., Clegg P., Hughes J. 1988, MNRAS, 235, 441

\refb Herbig G. H., Bell K. R. 1988, Lick Obs. Bull., No. 1111


\refb Hillenbrand L. A., Strom S. E., Vrba F. J., Keene J. 1992, ApJ,
397, 613

\refb Ivezi\'c \v{Z}., Becker R. H., Blanton M., Fan X. et al. 2002, in
{\it AGN Surveys} (IAU Colloq. 184), ASP Conf. Ser., 284, p. 137

\refb Jim\'enez-Esteban F. M., Agudo-M\'erida L., Engels D.,
Garcia-Lario P. 2005, A\&A, 431, 779

\refb Jim\'enez-Esteban F. M., Garcia-Lario P., Engels D., Perea
Calderon J.V. 2006, A\&A, 446, 773

\refb Karr J. L., Martin P. G. 2003a, ApJ, 595, 880

\refb Karr J. L., Martin P. G. 2003b, ApJ, 595, 900

\refb Kenyon S. J., Hartmann L. 1995, ApJS, 101, 117

\refb Kenyon S. J., Hartmann L. W., Strom K. M., Strom S. E. 1990, AJ,
99, 869

\refb Kerton C. R., Brunt C. M. 2003, A\&A, 399, 1083

\refb Kerton C. R., Martin P. G., Johnstone D., Ballantyne D. R. 2001,
ApJ, 552, 601

\refb Kwok S., Volk K., Bidelman W. P. 1997, ApJS, 112, 557

\refb Lada C. J. 1987, in {\it Star Forming Regions} (IAU Symp. 115),
eds. M. Peimbert \& J. Jugaku, Reidel Publ. Comp., Dordrecht, p.\,1

\refb Lada C. J., Adams F. C. 1992, ApJ, 393, 278

\refb Lewis B. M., Kopon D. A., Terzian Y. 2004, AJ, 127, 501

\refb Magnani L., Caillaut J.-P., Buchalter A., Beichman C. A. 1995,
ApJS, 95, 159

\refb Megeath S. T., Herter T., Beichman C., Gautier N., Hester J. J.,
Rayner J., Shupe D. 1996, A\&A, 307, 775

\refb Meyer M. R., Calvert N., Hillenbrand L. A. 1997, AJ, 114, 288

\refb Ogura K., Sugitani K., Pickles A. 2002, AJ, 123, 2597

\refb Ojha D. K., Ghosh S. K., Kulkarni V. et al. 2004, A\&A, 415, 1039

\refb Persson S. E., Campbell B. 1987, AJ, 94, 416

\refb Persson S. E., Campbell B. 1988, AJ, 96, 1019

\refb Robitaille T. P., Whitney B. A., Indebetouw R., Wood K., Denzmore
P. 2006, ApJS, 167, 256

\refb Robitaille T. P., Whitney B. A., Indebetouw R., Wood K. 2007,
ApJS, 169, 328

\refb Ruch G. T., Jones T. J., Woodward C. E., Polomski E. F., Gehrz R.
D., Megeath S. T. 2007, ApJ, 654, 338

\refb Saito H., Saito M., Sunada K., Yonekura Y. 2007, ApJ, 659, 459

\refb Saurer W., Seeberger R., Weinberger R. 1997, AAS, 126, 247

\refb Sevenster M. N. 2002, AJ, 123, 2772

\refb Skrutskie M. F., Cutri R. M., Stiening R., Weinberg M. D. et al.
2006, AJ, 131, 1163

\refb Snell R. L., Scoville N. Z., Sanders D. B., Erickson N. R.
1984, ApJ, 284, 176

\refb Strai\v{z}ys V. 1992, {\it Multicolor Stellar Photometry}, Pachart
Publ.  House, Tucson, Arizona

\refb Strai\v{z}ys V., Laugalys V. 2007, Baltic Astronomy, 16, 167
(Paper I)

\refb Takata T., Yamada T., Saito M. et al. 1994, A\&AS, 104, 529

\refb van der Veen W.\,E.\,C.,\,J., Habing H. J. 1988, A\&A, 194, 125

\refb Walker H. J., Cohen M. 1988, AJ, 95, 1801

\refb Walker H. J., Cohen M., Volk K., Wainscoat R. J., Schwartz D. E.
1989, AJ, 98, 2163

\refb Whitelock P. A., Feast M. W., Marang F., Groenewegen M.\,A.\,T.
2006, MNRAS, 369, 751

\refb Whitelock P., Marang F., Feast M. 2000, MNRAS, 319, 728

\refb Whitelock P., Menzies J., Feast M. et al. 1994, MNRAS, 267, 711

\refb Whitney B. A., Wood K., Bjorkman J. E., Wolff M. J.. 2003a, ApJ,
591, 1049

\refb Whitney B. A., Wood K., Bjorkman J. E., Cohen M. 2003b, ApJ, 598,
1079

\refb Whitney B. A., Indebetouw R., Bjorkman J. E., Wood K. 2004, ApJ,
617, 1177

\refb Wouterloot J.\,G.\,A., Brand J. 1989, A\&AS, 80, 149

\refb Wouterloot J.\,G.\,A., Brand J., Fiegle K. 1993, A\&AS, 98, 589

\vskip4mm

{\bf Errata.} In the printed version of the description of Figure 1
(Section 2) and in Figure 1 caption the blue and green line colors
should be interchanged. In the arXiv version the text is corrected.

\end{document}